\documentclass[letterpaper, 10 pt, conference]{ieeeconf}
\IEEEoverridecommandlockouts 

\usepackage{cite}
\usepackage{amsmath}
\usepackage{amssymb}
\usepackage{bm}
\let\labelindent\relax
\usepackage{enumitem}
\usepackage{algorithm}
\usepackage{algpseudocode}
\usepackage{graphicx}
\usepackage[mathscr]{eucal}
\usepackage{xcolor}
\usepackage[hidelinks]{hyperref}
\usepackage{siunitx}
\graphicspath{{figures/}}

\newtheorem{proposition}{Proposition}
\newtheorem{lemma}{Lemma}
\newtheorem{definition}{Definition}
\newtheorem{theorem}{Theorem}
\newtheorem{assumption}{Assumption}
\newtheorem{corollary}{Corollary}
\newtheorem{remark}{Remark}

\newtheorem{property}{Property}

\DeclareMathOperator{\argmin}{argmin}

\DeclareMathOperator{\interior}{Int}

\DeclareRobustCommand{\intr}[1]{\interior \left(#1\right)}

\definecolor{gem-1}{rgb}{0.066,0.443,0.745}
\definecolor{gem-2}{rgb}{0.866,0.329,0}
\definecolor{gem-3}{rgb}{0.929,0.694,0.125}
\definecolor{gem-4}{rgb}{0.521,0.086,0.819}
\definecolor{gem-5}{rgb}{0.231,0.666,0.196}
\definecolor{gem-6}{rgb}{0.184,0.745,0.937}
\definecolor{gem-7}{rgb}{0.819,0.015,0.545}
\definecolor{gem-8}{rgb}{1,0.839,0.039}
\definecolor{gem-9}{rgb}{0.396,0.509,0.992}
\definecolor{gem-10}{rgb}{1,0.27,0.227}
\definecolor{gem-11}{rgb}{0,0.639,0.639}
\definecolor{gem-12}{rgb}{0.796,0.517,0.364}

\newcommand{\mat}[1]{\begin{bmatrix} #1 \end{bmatrix}}

\newcommand{\N}{\mathbb{N}}

\newcommand{\R}{\mathbb{R}}

\newcommand{\cK}{\mathscr{K}}

\newcommand{\sB}{\mathcal{B}}
\newcommand{\sC}{\mathcal{C}}
\newcommand{\sD}{\mathcal{D}}

\newcommand{\sI}{\mathcal{I}}
\newcommand{\sJ}{\mathcal{J}}
\newcommand{\sK}{\mathcal{K}}

\newcommand{\sN}{\mathcal{N}}

\newcommand{\sS}{\mathcal{S}}
\newcommand{\sT}{\mathcal{T}}
\newcommand{\sU}{\mathcal{U}}
\newcommand{\sV}{\mathcal{V}}

\newcommand{\sX}{\mathcal{X}}

\usepackage{tikz}

\newcommand\copyrighttext{%
  \footnotesize \textcopyright 2026 IEEE. Personal use of this material is permitted.
  Permission from IEEE must be obtained for all other uses, in any current or future
  media, including reprinting/republishing this material for advertising or promotional
  purposes, creating new collective works, for resale or redistribution to servers or
  lists, or reuse of any copyrighted component of this work in other works.}
\newcommand\copyrightnotice{%
\begin{tikzpicture}[remember picture,overlay]
\node[anchor=south,yshift=10pt] at (current page.south) 
  {\fbox{\parbox{\dimexpr\textwidth-\fboxsep-\fboxrule\relax}{\copyrighttext}}};
\end{tikzpicture}%
}

\begin{document}
\title{Using Dynamic Safety Margins as Control Barrier Functions}
\author{Victor Freire, and Marco M. Nicotra
\thanks{This research was supported by the NSF-CMMI Award \#2411667.}
\thanks{The authors are with the Department of Electrical, Computer \& Energy Engineering, University of Co\-lorado,  Boulder, CO 80309 USA (email: vifr9883@colorado.edu; marco.nicotra@colorado.edu).} \thanks{\emph{(Corresponding author: Victor Freire.)}}}

\maketitle
\copyrightnotice
\begin{abstract}
This paper presents an approach to design control barrier functions (CBFs) for arbitrary state and input constraints using tools from the reference governor literature. In particular, it is shown that dynamic safety margins (DSMs) are CBFs for an augmented system obtained by concatenating the state with a virtual reference. The proposed approach is agnostic to the relative degree and can handle multiple state and input constraints using the control-sharing property of CBFs. The construction of CBFs using Lyapunov-based DSMs is then investigated in further detail. Numerical simulations show that the method outperforms existing DSM-based approaches, while also guaranteeing safety and persistent feasibility of the associated optimization program.
\end{abstract}

\section{Introduction}
Control barrier functions (CBFs) have become a po\-pular tool for deriving optimization-based constrained control laws that are safe, easy to implement, and achieve good performance \cite{molnar2023safety, heshmati2024control, tayal2024control}. Despite their success, the widespread use of CBFs is limited by the absence of a systematic method to synthesize them for general classes of systems with arbitrary state and input constraints.

Given a state constraint set, it is usually straightforward to find a function of the state vector that is positive for constraint-admissible states, and negative otherwise. We call such functions candidate CBFs. They are commonly used in practice to design CBF-based controllers and, with tuning, can achieve great performance (e.g., \cite{heshmati2024control}). However, candidate CBFs lack many safety guarantees that valid CBFs provide. In particular, the CBF-based optimization program governing the control law may become infeasible when using candidate CBFs, which is unacceptable for safety-critical systems. Ultimately, CBFs are certificates of control invariance. Therefore, if the original state constraint set is not control invariant, it is impossible to find a valid CBF associated to it. In these cases, we must resort to finding CBFs for a, hopefully large, control invariant subset of the state constraint set. In this work, we use tools from the reference governor literature \cite{garone2017reference} to systematically design valid CBFs given arbitrary state and input constraints.

\begin{figure}[t]
    \centering
    \includegraphics[width=0.9\linewidth]{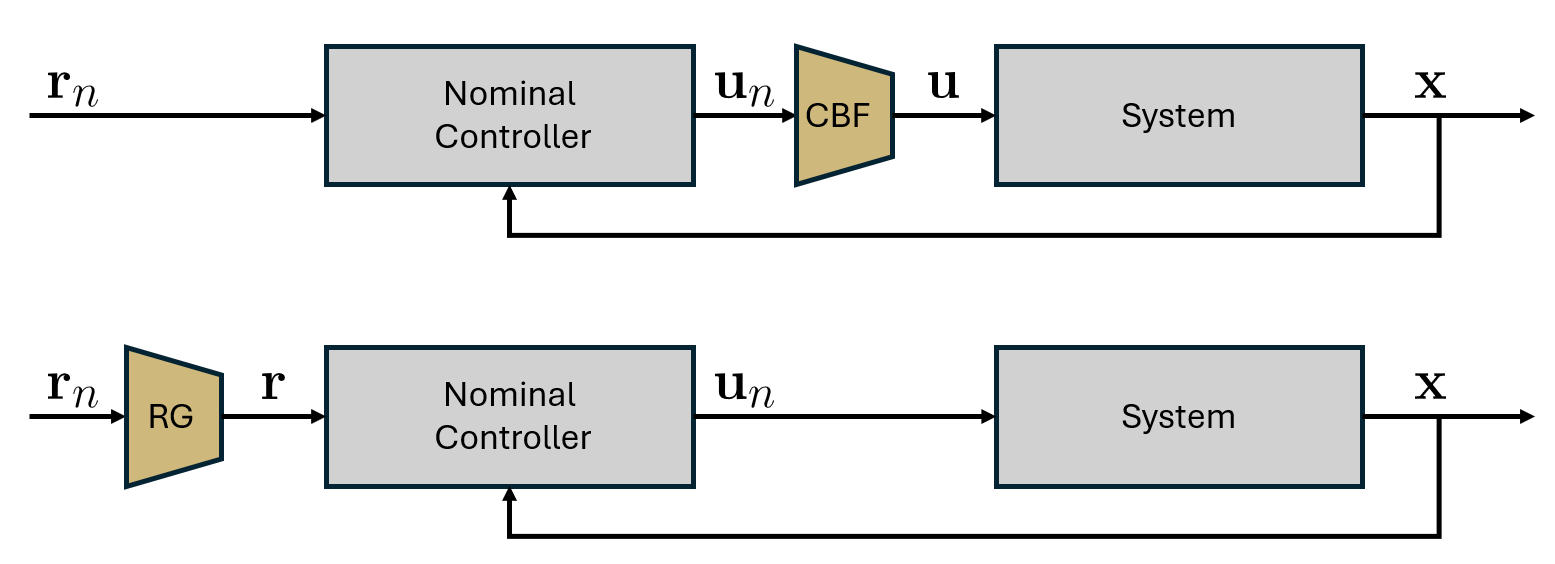}
    \caption{Comparison of CBF-based control (top) with RG-based control (bottom). The CBF approach filters a nominal input signal $\bm{u}_n$ to obtain a safe input $\bm{u}$. The RG approach a nominal reference signal $\bm{r}_n$ to obtain a safe virtual reference $\bm{r}$ for the nominal controller.}
    \label{fig:blk-diagram}
\end{figure}

\subsection{Related Works}
While the term control barrier function can be traced back to the early 2000's (see \cite{ames2019control} for a historical account), the modern definition that we consider in this work was first presented in \cite{ames2016control} under the name of zeroing CBFs. Since then, many works have studied different properties of CBFs and the systems they are used in: \cite{nguyen2016exponential, breeden2021high} address the relative degree of CBFs; \cite{xu2018constrained} defines control-sharing CBFs to enforce multiple constraints at once; \cite{agrawal2017discrete, zhang2022control} consider discrete-time and sampled-data systems, respectively; \cite{jankovic2018robust, buch2021robust} study robust CBFs.

Most relevant to the present manuscript, we review some works that study the design of CBFs and feasibility of the CBF-based optimization program.
In \cite{clark2024semi}, sum of squares (SOS) techniques are used to synthesize CBFs for semi-algebraic sets and polynomial systems using an alternating descent algorithm.
In \cite{dai2023convex}, the authors use SOS techniques to synthesize polynomial control Lyapunov functions, and then present an extension to construct CBFs.
In \cite{zeng2021decay}, the authors introduce a variable modifying the decay rate of the CBF to guarantee feasibility of the CBF-based program when the state is in the interior of the safe set even under input constraints. However, their approach provides no feasibility guarantee when the state is on the boundary of the safe set.
In \cite{choi2021robust}, the authors propose robust control barrier-value functions, which unify the Hamilton--Jacobi reachability and CBF methods, leading to viability kernel-sized invariant sets and CBF-like performances. This method, however, suffers from the curse of dimensionality.
In \cite{jang2024safe}, the authors use control Lyapunov functions (CLFs) to design a continuum of valid CBFs and use tools from hybrid systems theory to study the resulting safety properties.
In \cite{cortez2022safe}, the authors design CBFs for Euler--Lagrange systems. While the results are promising and the class of systems is relevant, the approach is  limited to box constraints.
In \cite{saveriano2019learning}, the authors learn CBFs from human demonstrations but they provide no guarantees that the learned CBFs are valid.
In \cite{dai2022learning}, given a valid CBF, the authors use learning to construct a CBF for a larger control invariant set using a combination of safe and unsafe trajectories of the system. Nonetheless, finding the initial, valid CBF remains an open question.
In \cite{chen2021backup}, the authors use a backup control policy to enlarge a small (but easy to find) control invariant set.
The enlarged control invariant set is then used to derive a backup CBF. These backup CBFs rarely have a closed form, which makes them difficult to implement.
In \cite{freire2023systematic}, the authors use maximal output admissible sets (MOASs) to design discrete-time CBFs for arbitrary state and input constraints. While the approach works well for linear systems, finding the MOAS for nonlinear systems remains 
an open question.

This work expands the underlying theme of \cite{chen2021backup, freire2023systematic}, where CBFs are obtained starting from a prestabilizing (or backup) controller. To do this, we adopt the notion of dynamic safety margin (DSM) from the explicit reference governor (ERG) framework \cite{nicotra2018explicit}. Given a fixed reference, a DSM measures the distance to constraint violation of a prestabilized system trajectory as it converges to the associated equilibrium point.
In \cite{li2023governor}, the authors define governor-parameterized barrier functions (PBFs), which are based on CBFs, and show that they are DSMs. While the results are limited to linear systems, their work is one of the few connections between the reference governor and CBF literatures. 
In contrast to their result, we show that DSMs are CBFs for an augmented system consisting of the concatenated state and reference vectors.
These works serve as motivation for exploring the connections between reference governor approaches like the command governor (CG) \cite{bemporad1997nonlinear}, which filters a nominal reference, and CBF-based control, which filters a nominal input. Both approaches ultimately modify a nominal, but potentially unsafe signal to guarantee closed-loop safety (see Fig. \ref{fig:blk-diagram}).

\subsection{Contributions}
The contributions of this manuscript are:
\begin{itemize}
  \item We leverage the control-sharing property presented in \cite{xu2018constrained} to define a natural generalization of traditional real-valued control barrier functions (CBFs) to vector-valued CBFs.
  \item We show that dynamic safety margins (DSMs) are CBFs for an augmented system consisting of the concatenation of the state vector and the reference of the prestabilizing controller used to construct the DSM. Additionally, we show that the relative degree of the constraints is irrelevant for DSMs by construction.
  \item We formulate the CBF safety filter in the augmented state-reference space and show that it is  persistently feasible. Moreover, the optimization problem is a quadratic program (QP) when the input constraints are polyhedral.
  \item We extend the Lyapunov-based DSMs presented in \cite{nicotra2018explicit} to inadmissible references, needed to use them as CBFs. Moreover, we show that any class $\cK$ function satisfies the CBF condition of Lyapunov-based DSMs.
  \item We present a comparison of our proposed approach with the explicit reference governor (ERG) \cite{nicotra2018explicit}, candidate CBFs and backup CBFs \cite{chen2021backup} in two nonlinear systems. The numerical simulations show that the proposed DSM-CBFs outperform the ERG, while boasting strict safety guarantees that are otherwise lost when candidate CBF-based control becomes infeasible. The examples also show that, while backup CBFs can achieve better performance than Lyapunov-based DSM-CBFs, their implementation is more computationally expensive due to its reliance on trajectory predictions.
\end{itemize}

The rest of the paper is organized as follows: Section \ref{sec:preliminaries} provides formal definitions for control invariance, CBFs, and DSMs; Section \ref{sec:DSM->CBF} proves that DSMs are CBFs, shows the relative degree of the constructed CBF is irrelevant for safety, formulates the CBF-based optimization problem, which is a quadratic program (QP) for polyhedral input constraints, and details some conditions that guarantee the DSM-CBF policy is locally Lipschitz continuous; Section \ref{sec:lyap-dsm} provides a systematic approach based on Lyapunov functions to construct DSMs and highlights some desirable properties that such DSMs have; Section \ref{sec:summary} summarizes the presented CBF synthesis approach and contextualizes it with respect to the existing reference governor literature; Section \ref{sec:examples} presents the numerical simulations and comparisons; Section \ref{sec:conclusion} concludes the paper.

\subsection{Notation}
Throughout the paper, only vector quantities are presented in bold font: $\bm{x} \in \R^n$. The notation $[\bm{x};~\bm{y}]$ denotes the vertical stacking of two vectors $\bm{x}$ and $\bm{y}$ into a column vector. A continuous function $\alpha:[0,\infty) \to [0,\infty)$ is said to belong to class $\cK$ if it is strictly increasing and $\alpha(0) = 0$.
The symbol $\subset$ denotes set inclusion, but not necessarily strict inclusion. Given a set $\sS$, we denote by $\intr{\sS}$ the interior of $\sS$ and by $\partial \sS$ the boundary of $\sS$ (i.e. $\partial \sS = \text{closure}(\sS) \setminus \intr{\sS}$). Given a subset $\sC \subset \R^n$ of the state space $\R^n$, we generally denote by $\tilde{\sC} \subset \R^n \times \R^l$ its counterpart in the augmented state-reference space. Let $h: \sD \to \R$ be continuously differentiable. We say $\bm{x} \in \sD$ is a \emph{regular point} of $h$ if $\partial h(\bm{x})/ \partial \bm{x} \neq 0$. We say $c \in \R$ is a \emph{regular value} of $h$ if every $\bm{x}\in \sD$ such that $h(\bm{x}) = c$ is a regular point.

\section{Preliminaries} \label{sec:preliminaries}
This section recalls the notion of control invariance, control barrier functions, and dynamic safety margins. Consider a control-affine system
\begin{equation}\label{eq:sys}
    \dot{\bm{x}} = f(\bm{x}) + g(\bm{x})\bm{u},
\end{equation}
where $f:\R^n \to\R^n$ and $g: \R^n \to \R^{n\times m}$ are locally Lipschitz continuous functions. The system is subject to state $\bm{x} \in \sX$ and input $\bm{u}\in\sU$ constraints, where $\sX \subset \R^n$ and $\sU \subset \R^m$ are closed sets.

\subsection{Control Invariance} \label{ssec:ctrl-inv}
Control invariance is a desirable property of state-space subsets that is closely related to safety. Roughly, a set is control invariant if there exists a suitable choice of inputs such that the trajectory of the closed-loop system remains in the set.
\begin{definition}[\!\!\cite{blanchini1999set}] \label{def:ci} A set $\sC \subset \R^n$ is \emph{control invariant} if, $\forall \bm{x}_0 \in \sC$, there exists an admissible, essentially bounded input signal $u:[0,\infty) \to \sU$ such that at least one of the trajectories of the closed-loop system \eqref{eq:sys}, with $x(0) = \bm{x}_0$, satisfies $x(t) \in \sC$ for all $t$ in its interval of existence $\sJ(\bm{x}_0) \subset [0,\infty)$.
\end{definition}
Note that this property is sometimes called weak control invariance \cite{blanchini1999set} or viability \cite{aubin1991viability} and that it doesn't require existence and uniqueness of the trajectory. Nonetheless, it will simplify the presentation of the upcoming results. Assumptions for the existence and uniqueness of solutions will be presented separately. Central to the upcoming discussion is Nagumo's theorem characterizing control invariance in terms of Bouligand's tangent cone.
\begin{theorem}[Nagumo {\cite{nagumo1942lage}}] \label{thm:nagumo}
   A closed set $\sC \subset \R^n$ is control invariant if and only if
  \begin{equation}
    \forall \bm{x} \in \sC, \quad \exists \bm{u} \in \sU, \quad f(\bm{x}) + g(\bm{x}) \bm{u} \in \sT_{\sC}(\bm{x}),
  \end{equation}
  where $\sT_{\sC}(\bm{x})$ is the tangent cone to $\sC$ in $\bm{x}$ \cite[Definition 3.1]{blanchini1999set}.
\end{theorem}
While this characterization was groundbreaking, working with the tangent cone $\sT_{\sC}(\bm{x})$ can be cumbersome.  
Fortunately, when the set $\sC$ satisfies certain regularity and transversality conditions (see \cite{aubin2009set} for details), the tangent cone on the boundary is simplified.
\begin{property}[MFCQ \cite{mangasarian1967fritz}]
  Let $h: \sD \to \R^{n_c}$ be continuously differentiable and denote its channels by $h_i: \sD \to \R$. We say $h$ satisfies the MFCQ if
  \begin{equation}\label{eq:MFCQ}
    \forall \bm{x} \in \sD, \quad \exists \bm{z} \in \R^n, \quad \forall i \in \sI(\bm{x}), \quad \frac{\partial h_i}{\partial \bm{x}} \bm{z} > 0,      
  \end{equation}
  where $\sI(\bm{x}) = \{i \mid h_i(\bm{x}) = 0\}$ is the set of active constraints.
\end{property}
\begin{proposition}[\!\!{\cite[Proposition 4.3.7]{aubin2009set}}] \label{prop:contingent-cone-smooth-intersection}
  Let $h:\sD \to \R^{n_c}$ be continuously differentiable. If $h$ satisfies the MFCQ, then
  \begin{equation}
      \forall \bm{x} \in \partial \sC, \quad \sT_{\sC}(\bm{x}) = \left\{ \bm{z} \in \R^n \bigg| \min_{i \in \sI(\bm{x})} \frac{\partial h_i}{\partial \bm{x}}\bm{z} \geq 0 \right\},
  \end{equation}
  where $\sC = \{\bm{x} \in \sD \mid \min_{i \in \{1,\ldots,n_c\}} h_i(\bm{x}) \geq 0\}$.
\end{proposition}
Theorem \ref{thm:nagumo} and Proposition \ref{prop:contingent-cone-smooth-intersection} imply the following result.
\begin{theorem} \label{thm:nagumo-smooth-intersection}
  Let $h:\sD \to \R^{n_c}$ be a continuously differentiable function that satisfies the MFCQ. The set
  \begin{equation} \label{eq:setC}
    \sC = \left\{\bm{x} \in \sD \;\Big|\; \min_{i \in \{1,\ldots,n_c\}} h_i(\bm{x}) \geq 0\right\},
  \end{equation}
  is control invariant if and only if $\forall \bm{x} \in \partial \sC$, $\exists \bm{u} \in \sU$, such that
  \begin{equation} \label{eq:nagumo-smooth-intersection}
    \min_{i \in \sI(\bm{x})} \left[ L_fh_i(\bm{x}) + L_gh_i(\bm{x}) \bm{u} \geq 0\right],
  \end{equation}
  where $L_fh_i$ and $L_gh_i$ are the Lie derivatives of $h_i$ along $f$ and $g$, respectively.
\end{theorem}

Although much simpler than evaluating the tangent cone, the conditions of Theorem \ref{thm:nagumo-smooth-intersection} apply only on the boundary of the set $\partial \sC$, which makes implementing them difficult in practice. This motivated the development of control barrier functions, which allow us to extend these conditions to the entirety of the set $\sC$.

\subsection{Control Barrier Functions}\label{ssec: CBF}
Control barrier functions certify the control invariance of their superlevel set. Traditionally, CBFs are real-valued functions \cite{ames2016control} and, when multiple CBFs need to be imposed simultaneously, the control-sharing property presented in \cite{xu2018constrained} ensures their compatibility. In this work, we combine these ingredients to present a general, vector-valued CBF.
\begin{definition} \label{def:cbf}
  A continuously differentiable function $h: \sD \to \R^{n_c}$ is a \emph{control barrier function} (CBF) if there exists $\alpha \in \cK$ for which, $\forall \bm{x} \in \sC$, $\exists \bm{u} \in \sU$, such that
  \begin{equation} \label{eq:cbf-cond}
    \min_{i \in \{1,\ldots,n_c\}} \left[L_fh_i(\bm{x}) + L_gh_i(\bm{x})\bm{u} + \alpha\big(h_i(\bm{x})\big)\right] \geq 0,
  \end{equation}
  where $\sC$ is given in \eqref{eq:setC}.
\end{definition}
\smallskip
\begin{remark} \label{rem:unique-alpha}
  While we stated Definition \ref{def:cbf} using a unique class $\cK$ function $\alpha$ to simplify the presentation, it is equivalent to considering a different $\alpha_i \in \cK$ for each $h_i$. Tuning each $\alpha_i$ allows the user to modulate the performance of the closed-loop system while retaining the safety guarantees as long as \eqref{eq:cbf-cond} is satisfied with $\alpha \to \alpha_i$.
\end{remark}

With this definition, the conditions of Theorem \ref{thm:nagumo-smooth-intersection} can be extended to the entire set $\sC$, as shown in the next result.
\begin{theorem} \label{thm:cbf-ci}
  Let $h:\sD \to \R^{n_c}$ be continuously differentiable and satisfy the MFCQ. Consider the set $\sC$ defined as in \eqref{eq:setC} and the set-valued map $\sK(\bm{x}) = \{\bm{u} \in \sU \mid \eqref{eq:nagumo-smooth-intersection}\}$.
  \begin{enumerate}
    \item If the function $h$ is a CBF, then $\sC$ is control invariant.\label{thm:cbf->ci}
    \item Assume (a) $\sC$ is compact and (b) there exists a continuous selection $\mu: \sC \to \sU$ such that $\mu(\bm{x}) \in \sK(\bm{x})$. If $\sC$ is control invariant, then $h$ is a CBF. \label{thm:ci->cbf}
  \end{enumerate}
\end{theorem}
\begin{proof}
  Although the proof is based on \cite{ames2016control}, they showed the result for real-valued barrier functions. Our result concerns vector-valued control barrier functions.

  We begin by showing statement \ref{thm:cbf->ci}. Assume $h$ is a CBF and let $\alpha \in \cK$ satisfy the CBF definition. Let $\bm{x} \in \partial \sC$ be given and pick $\bm{u} \in \sU$ such that \eqref{eq:cbf-cond} holds. Let $i \in \sI(\bm{x})$ and note that $0 \leq L_fh_i(\bm{x}) + L_gh_i(\bm{x}) \bm{u} +   \alpha\big(h_i(\bm{x})\big) = L_fh_i(\bm{x}) + L_gh_i(\bm{x}) \bm{u} + \alpha(0) =   L_fh_i(\bm{x}) + L_gh_i(\bm{x}) \bm{u}$.  We conclude by Theorem \ref{thm:nagumo-smooth-intersection} that $\sC$ is control invariant.

  Next, we show statement \ref{thm:ci->cbf}. Let $\sC$ be control invariant and pick $i \in \{1,\ldots,n_c\}$. Define the set-valued map
  \begin{equation}
    \sC_i(c) = \{\bm{x} \in \sC \mid 0 \leq h_i(\bm{x}) \leq c\},
  \end{equation}
  and the value function $\hat{\alpha}_i: [0, \infty) \to \R$ by
  \begin{equation}
    \hat{\alpha}_i(c) = - \inf_{\bm{x} \in \sC_i(c)} \left[L_fh_i(\bm{x}) + L_gh_i(\bm{x}) \mu(\bm{x})\right].
  \end{equation}
  For any $c \geq 0$, the set $\sC_i(c) \subset \sC$ is compact. Next, we argue the set-valued map $\sC_i$ is upper and lower hemicontinuous at $0$: Let $\{c_k\} \subset [0,\infty)$ be a sequence that converges to $0$ and let $\{\bm{x}_k\}\subset \sD$ be a sequence that converges to $\bm{x} \in \sD$ satisfying $\bm{x}_k \in \sC_i(c_k)$. Noting that $0 \leq h_i(\bm{x}_k) \leq c_k$ for all $k \in \N$, it follows by continuity of $h_i$ that $0 \leq h_i(\bm{x}) \leq 0$, and thus $\bm{x} \in \sC_i(0)$, concluding that $\sC_i $ is upper hemicontinuous at $0$; For lower hemicontinuity, pick any $\bm{x} \in \sC_i(0)$ and any sequence $\{c_k\} \subset [0,\infty)$ that converges to $0$. Clearly, $\bm{x} \in \sC_i(c_k)$ for all $k \in \N$, so we can create the trivial sequence $\{\bm{x}_k\} \subset \sD$ with $\bm{x}_k = \bm{x}$ to conclude that $\sC_i$ is lower hemicontinuous at $0$. With this, it follows by Berge's maximum theorem \cite{berge1963topological} that the value function $\hat{\alpha}_i$ is continuous at $0$. By construction, $\hat{\alpha}_i$ is non-decreasing. Next, let $\bm{x} \in \sC_i(0) \subset \sC$ be given and note that since $\mu(\bm{x}) \in \sK(\bm{x})$, $L_fh_i(\bm{x}) + L_gh_i(\bm{x}) \mu(\bm{x}) \geq 0$, and it follows that $\hat{\alpha}_i(0) \leq 0$. There always exists $\alpha_i \in \cK$ that upper-bounds $\hat{\alpha}_i$. Since $i \in \{1,\ldots, n_c\}$ was arbitrary, let us define  $\alpha(c) = \max_{i \in \{1,\ldots,n_c\}} \alpha_i(c)$ and note that $\alpha \in \cK$.
  Finally, let $\bm{x} \in \sC$ be given and recall $\mu(\bm{x}) \in \sU$. For any $i \in \{1,\ldots,n_c\}$,
  \begin{align*}
    \dot{h}_i\big(\bm{x}, \mu(\bm{x})\big) &=    L_fh_i(\bm{x}) + L_gh_i(\bm{x}) \mu(\bm{x})\\
    & \geq \inf_{\bm{x}_0 \in
    \sC_i\big(h_i(\bm{x})\big)} \left[L_fh_i(\bm{x}_0) + L_gh_i(\bm{x}_0)
    \mu(\bm{x}_0)\right] \\
    &\geq -\hat{\alpha}_i\big(h_i(\bm{x})\big) \\
    &\geq -\alpha_i\big(h_i(\bm{x})\big) \\
    &\geq -\alpha\big(h_i(\bm{x})\big).
  \end{align*}
  We conclude that $h$ is a CBF.
\end{proof}
\begin{remark}
  When $\sC$ is control invariant, the set $\sK(\bm{x}) \neq \emptyset$ for every $\bm{x} \in \sC$. Therefore there always exists a selection $\mu: \sC \to \sU$ such that $\mu(\bm{x}) \in \sK(\bm{x})$. Statement \ref{thm:ci->cbf} of Theorem \ref{thm:cbf-ci} requires at least one such selection be continuous. Moreover, in the special case that $n_c = 1$, the following statement also holds: \emph{Assume (a) $\sC$ is compact and (b) $\sU$ is compact. If $\sC$ is control invariant, then $h$ is a CBF}.
\end{remark}
\smallskip
CBFs can be used to design add-on modules that enforce constraint satisfaction by filtering the control input. Specifically, let $h:\sD \to \R^{n_c}$ be a CBF such that $\sC \subset \sX$ and consider the closed-loop system
\begin{equation} \label{eq:sys-cbf}
    \dot{\bm{x}} = f(\bm{x}) + g(\bm{x})\beta(\bm{x}),
\end{equation}
where $\beta: \sC \to \sU$ is the optimization-based control policy
\begin{equation} \label{eq:beta}
    \beta
    (\bm{x}) = \underset{\bm{u} \in \sK_\alpha(\bm{x})}\argmin \: \|\bm{u}-\kappa(\bm{x})\|^2,
\end{equation}
where $\kappa: \R^n \to \R^m$ is a nominal control policy with desirable closed-loop performance and $\sK_{\alpha}(\bm{x}) = \{\bm{u} \in \sU \mid \eqref{eq:cbf-cond}\}$. Since $h$ is a CBF, there exists $\alpha \in \cK$ such that the optimization problem \eqref{eq:beta} is always feasible, making $\beta$ well defined. If, in addition, the policy $\beta$ is locally Lipschitz continuous, there exists a unique solution $x(t)$ to the closed-loop system \eqref{eq:sys-cbf}, and it satisfies $x(t) \in \sC \subset \sX$. In upcoming sections, we provide conditions under which $\beta$ is locally Lipschitz continuous.

As noted in the introduction, finding a CBF for arbitrary constraint sets $\sX \subset \R^n$ is a challenging task. In fact, Theorem \ref{thm:cbf-ci} states that a CBF $h:\sD \to \R^{n_c}$ with superlevel set $\sX$ exists only if $\sX$ is control invariant, which is generally not the case. In spite of this, many practitioners insist on imposing $\sC = \sX$ to construct a ``candidate'' CBF (i.e. a continuously differentiable function $h:\sD \to \R^{n_c}$ that satisfies $\sX = \sC$ with $\sC$ as in \eqref{eq:setC}) and then tuning a class $\cK$ function $\alpha$ on a case-by-case basis. Although this approach can yield good performance, it suffers from the possibility that $\sK_{\alpha}(\bm{x}) = \emptyset$ for some $\bm{x}\in\sX$, which makes the controller $\beta$ ill-defined.

In this paper, we leverage results from the ERG framework to design valid CBFs that satisfy Definition \ref{def:cbf}.

\subsection{Dynamic Safety Margins}\label{ssec: DSM}
Dynamic safety margins are functions that quantify the distance to constraint violation for a prestabilized system subject to a constant reference. In order to use them, we must make the following assumptions about system \eqref{eq:sys}. 
\begin{assumption}\label{ass:stab}
System \eqref{eq:sys} admits a simply connected equilibrium manifold parameterized by continuous functions $\bar{x}:\R^l \to \R^n$ and $\bar{u}: \R^l \to \R^m$ such that 
\begin{equation}
    \forall \bm{v} \in \R^l, \quad f\big(\bar{x}(\bm{v})\big) + g\big(\bar{x}(\bm{v})\big) \bar{u}(\bm{v}) = 0.
\end{equation}
Moreover, every equilibrium point $\bar x(\bm{v})$ is stabilizable. 
\end{assumption}
\smallskip
Assumption \ref{ass:stab} is standard in the reference governor literature: it applies to any system that admits a connected path of stabilizable equilibrium points. As noted in \cite{garone2017reference}, this encompasses a wide range of meaningful systems and applications.

The vector $\bm{v} \in \R^l$ is hereafter called a ``reference'' for the system because it identifies a specific equilibrium point. The set of steady-state admissible references is 
\begin{equation}\label{eq:ss-adm}
    \sV = \{\bm{v} \in \R^l \mid \bar{x}(\bm{v}) \in \sX, ~\bar{u} (\bm{v}) \in \sU\}.
\end{equation}
Note that, since $\bar{x}$, $\bar{u}$ are continuous  and $\sX$, $\sU$ are closed sets, $\sV$ is closed in $\R^l$.
\begin{assumption} \label{ass:adm-eq}
    The set $\sV \!\subset\! \R^l$ is not empty, i.e., system \eqref{eq:sys} features at least one steady-state admissible equilibrium point.
\end{assumption}
\smallskip

In essence, Assumption \ref{ass:stab} establishes a continuous mapping from references to stabilizable equilibria, and Assumption \ref{ass:adm-eq} states that at least one of these equilibria is safe. By Assumption \ref{ass:stab}, let $\pi:\R^n \times \R^l \to \R^m$, locally Lipschitz continuous on $\R^n$ and continuous on $\R^l$, be a prestabilizing control law such that $\pi\big(\bar{x}(\bm{v}),\bm{v}\big) = \bar{u}(\bm{v})$, and $\bar{x}(\bm{v})$ is an asymptotically stable equilibrium point of the prestabilized system
\begin{equation}
    \dot{\bm{x}} = f_{\pi}(\bm{x}, \bm{v}) \triangleq f(\bm{x}) + g(\bm{x}) \pi(\bm{x},\bm{v}).
\end{equation}
Given a reference $\bm{v} \in \R^l$, the prestabilized dynamics $f_{\pi}$ ensure the existence of an open set $\sD_{\bm{v}} \subset \R^n$ that satisfies $\bar{x}(\bm{v}) \in \sD_{\bm{v}}$ and is such that
\begin{equation}
  \bm{x}(0) \in\sD_{\bm{v}}  \quad \implies  \quad \lim_{t \to \infty} \bm{x}(t) = \bar{x}(\bm{v}).
\end{equation}
The set $\sD_{\bm{v}}$ is a region of attraction for $\bar{x}(\bm{v})$.
Note that, for each reference $\bm{v} \in \R^l$, the input constraint set $\sU$ induces state constraints on the prestabilized system. This allows us to define the reference-dependent state constraint set 
\begin{equation}\label{eq:Xv}
  \sX_{\bm{v}} = \{\bm{x} \in \sX \mid \pi(\bm x,\bm v)\in\sU\}. 
\end{equation}
By continuity of $\pi$ on $\R^l$, it follows that $\sX_{\bm{v}}$ is closed in $\R^n$.

Let us now define the state-reference space $\R^n \times \R^l$ obtained by concatenating the state and reference vectors $[\bm{x};~ \bm{v}]$.
\begin{assumption} \label{ass:open-attraction}
    The state-reference region of attraction set
    \begin{equation} \label{eq:DxV}
        \tilde{\sD} \triangleq \{(\bm{x}, \bm{v}) \in \R^n \times \R^l \mid \bm{x} \in \sD_{\bm{v}}\},
    \end{equation}
    is open in $\R^n \times \R^l$.
\end{assumption}
\smallskip
\begin{lemma}
  The state-reference constraint set
  \begin{equation}
    \tilde{\sX} \triangleq \{(\bm{x}, \bm{v}) \in \R^n \times \R^l \mid \bm{x} \in \sX_{\bm{v}}, ~ \bm{v} \in \sV\},
  \end{equation}
  is closed in $\R^n \times \R^l$.
\end{lemma}
\begin{proof}
  Let $\{(\bm{x}_n, \bm{v}_n)\} \subset \tilde{\sX}$ be a sequence that converges to $(\bm{x}, \bm{v}) \in \R^n \times \R^l$. Since $\{\bm{x}_n\} \subset \sX$ and $\{\bm{v}_n\} \subset \sV$, and $\sX$, $\sV$ are closed, it follows that $\bm{x} \in \sX$ and $\bm{v} \in \sV$. Define the sequence $\{\pi(\bm{x}_n, \bm{v}_n)\} \subset \sU$ and note that by continuity of $\pi$, it converges to $\pi(\bm{x}, \bm{v}) \in \sU$ because $\sU$ is closed. With this, we have that $\bm{x} \in \sX_{\bm{v}}$ and it follows that $(\bm{x}, \bm{v}) \in \tilde{\sX}$.
\end{proof}
The following DSM definition satisfies the one proposed in \cite[Definition 1]{nicotra2018explicit}, although it makes stronger assumptions by requiring a) differentiability rather than continuity and b) invariance rather than strong returnability.

\begin{definition}\label{def:dsm}
  Let $\Delta: \tilde{\sD} \to \R^{n_c}$ be continuously differentiable and define the set $\tilde{\sC} = \{(\bm{x}, \bm{v}) \in \tilde{\sD} \mid \Delta(\bm{x}, \bm{v}) \geq 0\}$. The function $\Delta$ is called a \emph{dynamic safety margin} (DSM) if the following conditions hold
  \begin{subequations}
    \begin{align}
        & \tilde{\sC} \subset \tilde{\sX}, \label{eq:dsm-safe}\\
        &\tilde{\sC} ~ \text{is compact in} ~ \R^n \times \R^l, \label{eq:dsm-compact}\\
        & \Delta_i(\bm{x}, \bm{v}) = 0 ~ \implies ~ \frac{\partial \Delta_i}{\partial \bm{x}} f_{\pi}(\bm{x}, \bm{v}) \geq 0, \label{eq:dsm-invariant}
    \end{align}
  \end{subequations}
  where the last implication holds for all $i \in \{1,\ldots, n_c\}$.
 \end{definition}

DSMs are used to design add-on modules (ERGs) that enforce constraint satisfaction by filtering the reference of the prestabilized system. 
Specifically, let us augment the prestabilized system with reference dynamics to obtain
\begin{equation} \label{eq:sys-erg}
    \mat{\dot{\bm{x}}\\ \dot{\bm{v}}} = \mat{f(\bm{x}) + g(\bm{x}) \pi(\bm{x}, \bm{v})  \\
    \min_{i \in \{1,\ldots,n_c\}}\Delta_i(\bm{x}, \bm{v})\rho(\bm{v}, \bm{r})},
\end{equation}
where $\rho: \R^l \times \R^l \to \R^l$ is any locally Lipschitz continuous function such that $\bm{r} \in \R^l$ is an asymptotically stable equilibrium point of the system $\dot{\bm{v}}(t) = \rho\big(\bm{v}(t),\bm{r}\big)$. In the ERG literature, $\rho$ is called a \emph{navigation field} \cite[Definition 2]{nicotra2018explicit}. System \eqref{eq:sys-erg} ensures that, if $\big(x(0), v(0)\big) \in \tilde{\sC}$, then $\big(x(t), v(t)\big) \in \tilde{\sC} \subset \tilde{\sX}$ at all times \cite{nicotra2018explicit}. The ERG literature provides systematic tools for computing DSMs for various classes of systems \cite{nicotra2015control, nicotra2018explicit,garone2018explicit, nicotra2018delay,li2020fast, cotorruelo2021reference}. The following section shows how these tools can be used to construct valid CBFs.

\section{From Dynamic Safety Margins to Control Barrier Functions} \label{sec:DSM->CBF}
This section explores the connections between DSMs and CBFs, formulates a safe, optimization-based control policy and provides conditions for this policy to be locally Lipschitz continuous.

\subsection{DSMs are CBFs}
Based on the intuition that both CBFs and DSMs can be used to design add-on modules for constraint handling, the following theorem states that a dynamic safety margin is, in fact, a CBF for an augmented system consisting of the concatenation of $\bm{x} \in \R^n$ and $\bm{v}\in \R^l$
\begin{equation} \label{eq:sys-aug}
    \mat{\dot{\bm{x}} \\ \dot{\bm{v}}} = \mat{f(\bm{x}) + g(\bm{x})\bm{u} \\ \bm{w}},
  \end{equation}
where $(\bm{u}, \bm{w}) \in \R^m \times \R^l$ are the augmented inputs.
\begin{theorem}[\bf{DSMs are CBFs}] \label{thm:dsm-cbf}
  If $\Delta: \tilde{\sD} \to \R^{n_c}$ is a DSM, then $\Delta$ is a CBF with respect to the augmented system \eqref{eq:sys-aug}.
\end{theorem}
\begin{proof}
  Define the set-valued maps
  \begin{equation}
    \tilde{\sC}_i(c) = \{(\bm{x}, \bm{v}) \in \tilde{\sC} \mid 0 \leq \Delta_i(\bm{x}, \bm{v}) \leq c\},
  \end{equation}
  and the functions $\hat{\alpha}_i: [0,\infty) \to \R$ by
  \begin{equation}
    \hat{\alpha}_i(c) = - \inf_{(\bm{x},\bm{v}) \in \tilde{\sC}_i(c)} \frac{\partial \Delta_i}{\partial \bm{x}} f_{\pi}(\bm{x}, \bm{v}),
  \end{equation}
  where $i \in \{1,\ldots,n_c\}$. Note that for any $c \in [0,\infty)$, the set $\tilde{\sC}_i(c) \subset \tilde{\sC}$ is compact. It can also be shown (see the proof of Theorem \ref{thm:cbf-ci}) that the set-valued maps $\tilde{\sC}_i$ are both upper and lower hemicontinuous at 0. Thus, it follows by Berge's maximum theorem \cite{berge1963topological} that each $\hat{\alpha}_i$ is continuous at 0. Moreover, it is nondecreasing because for any $c_1 \leq c_2$, $\tilde{\sC}_i(c_1) \subset \tilde{\sC}_i(c_2)$. Next, note that for any $(\bm{x}, \bm{v}) \in \tilde{\sC}_i(0)$, $\Delta_i(\bm{x}, \bm{v}) = 0$, and it follows by \eqref{eq:dsm-invariant} that
  \begin{equation*}
      \frac{\partial \Delta_i}{\partial \bm{x}} f_{\pi}(\bm{x}, \bm{v}) \geq 0.
  \end{equation*}
  Therefore,
  \begin{equation}
    \hat{\alpha}_i(0) = -\inf_{(\bm{x}, \bm{v}) \in \tilde{\sC}_i(0)} \frac{\partial \Delta_i}{\partial \bm{x}} f_{\pi}(\bm{x}, \bm{v}) \leq 0.
  \end{equation}
  There always exists $\alpha_i \in \cK$ such that $\forall c \in [0,\infty)$, $\hat{\alpha}_i(c) \leq \alpha_i(c)$. Define $\alpha(c) = \max_{i \in \{1,\ldots,n_c\}} \alpha_i(c)$ and note that $\alpha \in \cK$.
  Let $(\bm{x}, \bm{v}) \in \tilde{\sC}$ and $i \in \{1,\ldots,n_c\}$ be given and note that since $\bm{x} \in \sX_{\bm{v}}$, it follows that $\pi(\bm{x},\bm{v}) \in \sU$. With this,
  \begin{align*}
    \dot{\Delta}_i\big(\bm{x}, \bm{v}, \pi(\bm{x}, \bm{v}), 0\big) &= \frac{\partial \Delta_i}{\partial \bm{x}} f_{\pi}(\bm{x}, \bm{v}) + \frac{\partial \Delta_i}{\partial \bm{v}}0\\
    &\geq \inf_{(\bm{x}, \bm{v}) \in \tilde{\sC}_i(\Delta_i(\bm{x}, \bm{v}))} \frac{\partial \Delta_i}{\partial \bm{x}} f_{\pi}(\bm{x}, \bm{v}) \\
    &\geq -\hat{\alpha}_i\big(\Delta_i(\bm{x}, \bm{v})\big)\\
    &\geq -\alpha_i\big(\Delta_i(\bm{x}, \bm{v})\big)\\
    &\geq -\alpha\big(\Delta_i(\bm{x}, \bm{v})\big).
  \end{align*}
  We have shown that $\forall (\bm{x}, \bm{v}) \in \tilde{\sC}$, $(\bm{u},\bm{w}) = \big(\pi(\bm{x}, \bm{v}), 0\big) \in \sU \times \R^l$ is such that
  \begin{equation} \label{eq:dsm-cond}
    \min_{i \in \{1, \ldots, n_c\}} \left[\dot{\Delta}_i (\bm{x}, \bm{v}, \bm{u}, \bm{w}) + \alpha\big(\Delta_i(\bm{x}, \bm{v})\big)\right] \geq 0.
  \end{equation}
  We conclude that $\Delta$ is a CBF.
\end{proof}
Let us define the set of viable inputs at $(\bm{x},\bm{v})$ for DSM-CBFs
\begin{equation}
  \tilde{\sK}_{\alpha}(\bm{x}, \bm{v}) = \{(\bm{u},\bm{w})\in\sU\times \R^l \mid \eqref{eq:dsm-cond}\} \label{eq:dsm-adm-set}.
\end{equation}
The following result follows from Theorem \ref{thm:cbf-ci} and Theorem \ref{thm:dsm-cbf}.
\begin{corollary} \label{cor:dsm-ci}
  Let $\Delta: \tilde{\sD} \to \R^{n_c}$ be a DSM. If $\Delta$ satisfies the MFCQ, then the set $\tilde{\sC} \subset \tilde{\sX} \subset \sX \times \sV$ is control invariant.
\end{corollary}

A significant concern in the CBF literature arises when the term $L_gh(\bm{x}) = 0$ for some $\bm{x} \in \sC$. This means $h$ lacks uniform relative degree 1 over $\sC$, and implies a loss of controllability that can be catastrophic when $L_fh(\bm{x}) < -\alpha\big(h(\bm{x})\big)$. Although many relevant works address this issue (e.g. \cite{nguyen2016exponential, breeden2021high}), the following corollary shows an additional desirable property of DSM-CBFs under total controllability loss.

\begin{corollary} \label{cor:rel-deg-irrelevant}
  Let $\Delta: \tilde{\sD} \to \R^{n_c}$ be a DSM and let $\alpha \in \cK$ be such that \eqref{eq:dsm-cond} holds. If $(\bm{x},\bm{v}) \in \tilde{\sC}$ is such that
  \begin{equation}
  \mat{\frac{\partial \Delta}{\partial \bm{x}}g(\bm{x}) & \frac{\partial \Delta}{\partial \bm{v}}} = 0,
  \end{equation}
  then, $\tilde{\sK}_{\alpha}(\bm{x},\bm{v}) = \sU \times \R^l$.
\end{corollary}
\begin{proof}
  Note that since $\big(\pi(\bm{x}, \bm{v}), 0\big) \in \tilde{\sK}_{\alpha}(\bm{x}, \bm{v})$, it follows that $\forall i \in \{1,\ldots,n_c\}$,
  \begin{align*}
    \dot{\Delta}_i\big(\bm{x},\! \bm{v},\! \pi(\bm{x},\! \bm{v}),\! 0\big) \!&=\! \frac{\partial \Delta_i}{\partial \bm{x}}f(\bm{x}) \!+\! \mat{\frac{\partial \Delta_i}{\partial \bm{x}}g(\bm{x}) \!& \! \frac{\partial \Delta_i}{\partial \bm{v}}} \! \mat{\pi(\bm{x},\bm{v}) \\ 0}\\
    &= \frac{\partial \Delta_i}{\partial \bm{x}}f(\bm{x}) \geq -\alpha\big(\Delta_i(\bm{x},\bm{v})\big).
  \end{align*}
  Therefore, constraint \eqref{eq:dsm-cond} is trivially satisfied at $(\bm{x}, \bm{v})$ for any augmented input $(\bm{u}, \bm{w})$. We conclude that $\tilde{\sK}_{\alpha}(\bm{x},\bm{v}) = \sU \times \R^l$.
\end{proof}

\subsection{DSM-CBF Optimization-based Policy}
Recalling Theorem \ref{thm:dsm-cbf}, it is possible to reformulate the CBF policy $\beta$ for the augmented system $(\bm{x}, \bm{v})$. To do so, let $\bm{r} \in \R^l$ be a desired reference, let $\kappa:\R^n \times \R^l \to \R^m$ be a nominal control policy with desirable closed-loop performance, and let $\rho: \R^l \times  \R^l \to \R^l$ be a navigation field. The following theorem formulates the DSM-CBF-based policy $\tilde{\beta}$ for the augmented system \eqref{eq:sys-aug}.
\begin{theorem}[\bf{DSM-CBF policy}] \label{thm:dsm-qp}
    Let $\eta > 0$ be a constant scalar. If $\Delta : \tilde{\sD} \to \R^{n_c}$ is a DSM, then there exists $\alpha \in \cK$ such that the optimization-based policy $\tilde{\beta}: \tilde{\sC} \to \sU \times \R^l$
    \begin{equation} \label{eq:beta-dsm}
      \tilde{\beta}(\bm{x},\bm{v}) \! = \!\!\!\! \underset{(\bm{u},\bm{w}) \in \tilde{\sK}_{\alpha}(\bm{x},\bm{v})}\argmin \|\bm{u} \!-\!\kappa(\bm{x}, \bm{r})\|^2 \! +  \eta \| \bm{w} \!-\! \rho(\bm{v}, \bm{r})\|^2,
    \end{equation}
    is feasible for all $(\bm{x},\bm{v}) \in \tilde{\sC}$. Moreover, if $\tilde{\beta}$ is locally Lipschitz continuous and $\Delta$ satisfies the MFCQ, then the trajectory of the augmented system \eqref{eq:sys-aug} under policy $(\bm{u}, \bm{w}) = \tilde{\beta}(\bm{x},\bm{v})$ exists, is unique, and  satisfies $x(t) \in \sX$ and $v(t) \in \sV$ for all $t \geq 0$, given $\big(x(0), v(0)\big) \in \tilde{\sC}$.
\end{theorem}
\begin{proof}
  Feasibility of the DSM-CBF policy follows by Theorem \ref{thm:dsm-cbf}. Any locally Lipschitz controller in $\tilde{\sK}_{\alpha}$ certifies the local (in time) existence and uniqueness of the solution to the closed-loop system \cite{khalil2002nonlinear}. Recalling Corollary \ref{cor:dsm-ci}, control invariance of $\tilde{\sC}$ implies there can be no finite-time escape instances and the existence and uniqueness holds for all $t \geq 0$. Safety follows from $\tilde{\sC} \subset \sX \times \sV$.
\end{proof}
\smallskip

\begin{remark}
    Convexity of the set $\tilde{\sK}_{\alpha}(\bm{x},\bm{v})$, and therefore problem \eqref{eq:beta-dsm}, is linked to convexity of $\sU$. In particular, if the input constraint set $\sU$ is polyhedral, the optimization problem \eqref{eq:beta-dsm} becomes a quadratic program (QP) for all $(\bm{x},\bm{v}) \in \tilde{\sC}$, which can be solved efficiently.
\end{remark}
\begin{remark} \label{rem:eta}
  The positive constant $\eta > 0$ makes the objective function of the augmented CBF-program \eqref{eq:beta-dsm} strictly convex. Moreover, we observed in practice that as $\eta \to \infty$, the system behavior approaches that of the ERG, which tends to have worse performance than can be achieved with  $\eta \to 0$.
\end{remark}

\begin{remark}
    If $\sV$ is convex, a suitable navigation field is $\rho(\bm{v}, \bm{r}) = \bm{r} - \bm{v}$. Otherwise, please refer to works like \cite{rimon1991construction}.
\end{remark}

\begin{remark} \label{rem:tuning-pi}
    Note that the nominal controller $\kappa$ and the prestabilizing controller $\pi$ can be different. In practice, we observed that mild prestabilizing controllers $\pi$ lead to larger $\tilde{\sK}_{\alpha}$, which is the feasible region of \eqref{eq:beta-dsm}, allowing the safe input to be closer to the nominal target $\kappa$.
\end{remark}

\subsection{Existence and Uniqueness of Solutions}
Given the prevalence of optimization-based policies in constrained control approaches such as MPC and CBFs, numerous works have studied their regularity properties \cite{morris2013sufficient,jankovic2018robust,mestres2025regularity}.

As stated in Theorem \ref{thm:dsm-qp}, given any initial condition $(\bm{x}, \bm{v}) \in \tilde{\sC},$ the optimization-based policy $\tilde{\beta}$ for DSM-CBFs, defined in \eqref{eq:beta-dsm}, is always feasible. If, additionally, it is locally Lipschitz continuous, then the trajectory for the closed-loop system
\begin{equation} \label{eq:sys-dsmcbf}
  \mat{\dot{\bm{x}} \\ \dot{\bm{v}}} = \mat{f(\bm{x}) \\ 0} + \mat{g(\bm{x}) & 0 \\ 0& I}\beta(\bm{x}, \bm{v}),
\end{equation}
exists and is unique.
Based on \cite{mestres2025regularity}, we can formulate a reduced set of sufficient conditions that guarantee $\beta$ is locally Lipschitz continuous.

\begin{theorem}[\!\!{\cite[Theorem 5.2]{mestres2025regularity}}] \label{thm:lipschitz}
  Let $\Delta: \tilde{\sD} \to \R^{n_c}$ be a DSM, let $\rho$ be a navigation field and let $\eta > 0$. Assume the input constraints are polyhedral $\sU = \{\bm{u} \in \R^m \mid M\bm{u} \leq \bm{b}\}$, and the functions $\pi(\cdot, \bm{r})$ and $\rho(\cdot, \bm{r})$ are continuously differentiable.  If these constraint qualifications (CQs) hold
  \begin{enumerate}[label=\alph*)]
    \item (Slater CQ) There exists $\alpha \in \cK$ such that for all $(\bm{x}, \bm{v}) \in \tilde{\sC}$, $\intr{\tilde{\sK}_{\alpha}(\bm{x}, \bm{v})} \neq \emptyset$;
    \item (Constant Rank CQ) There exists a neighborhood of every pair $(\bm{x}, \bm{v}) \in \tilde{\sC}$ wherein, for any subset $\sJ \!\subset\! \sI(\bm{x},\bm{v})$, where $\sI(\bm{x}, \bm{v})$ is the set of active constraints, the number of linearly independent vectors in 
    \begin{equation*}
    \left\{\mat{\frac{\partial \Delta_i}{\partial \bm{x}}g(\bm{x}) & \frac{\partial \Delta_i}{\partial \bm{v}}}, i \in \sJ\right\}
    \end{equation*}
    remains constant;
  \end{enumerate}
  then the optimization-based policy $\beta$ is locally Lipschitz continuous and the
  solution to the closed-loop system \eqref{eq:sys-dsmcbf} exists and is unique.
\end{theorem}

\begin{remark} \label{rem:lipschitz}
  In the special case that $n_c = 1$, both Slater and Constant Rank CQs hold if $\Delta$ has uniform relative degree $1$ over $\tilde{\sC}$ and $\pi(\bm{x}, \bm{v}) \in \intr{\sU}$  for all $(\bm{x},\bm{v}) \in \tilde{\sC}$.
\end{remark}

\section{Lyapunov-based Dynamic Safety Margins}\label{sec:lyap-dsm}
This section shows that the Lyapunov-based DSMs in \cite{nicotra2018explicit, nicotra2015control, garone2018explicit, nicotra2018delay} can be used within the context of this paper. For trajectory-based DSMs, refer to \cite{freire2026designing} instead, which can be regarded as a generalization of backup CBFs \cite{chen2021backup} when using dynamic backup policies.
\begin{definition} \label{def:lyap}
  A continuously differentiable function $V: \R^n \times \R^l \to \R$ is a \emph{reference-dependent Lyapunov function} for the prestabilized system $f_{\pi}$ if there exist functions $\gamma_1, \gamma_2 \in \cK$ and $\forall \bm{v} \in \R^l$, there exists a neighborhood of $\bar{x}(\bm{v})$, denoted $\sD_{\bm{v}} \subset \R^n$, wherein
  \begin{subequations}\label{eq:lyap}
    \begin{align}
      \gamma_1\big(\|\bm{x} - \bar{x}(\bm{v})\|\big) &\leq V(\bm{x}, \bm{v}) \leq
      \gamma_2\big(\|\bm{x} - \bar{x}(\bm{v})\|\big), \label{eq:lyap-bounded} \\
      \frac{\partial V}{\partial \bm{x}}f_{\pi}(\bm{x},\bm{v}) & \leq
      0.\label{eq:lyap-decreases}
    \end{align}
  \end{subequations}
\end{definition}
\medskip

\begin{remark}
  Note that the reference-dependent Lyapunov function $V$ needs not certify asymptotic stability of the equilibria of the prestabilized system $f_{\pi}$. That is, $\dot{V}(\bm{x}, \bm{v}) = 0$ is acceptable for $\bm{x} \neq \bar{x}(\bm{v})$. This allows for more general classes of Lyapunov functions to be considered and ultimately simplifies the design of the DSM.
\end{remark}

Additionally, we require $V$ has non-vanishing gradient near the equilibrium points with the following assumption.
\begin{assumption}\label{ass:lyap-regular}
  Given any $\bm{v} \in \R^l$, $\bm{x} \in \sD_{\bm{v}}\setminus \{\bar{x}(\bm{v})\}$ is a regular point of $V(\cdot, \bm{v}): \R^n \to \R$.
\end{assumption}
Recalling the reference-dependent state constraint set $\sX_{\bm{v}}$ defined in \eqref{eq:Xv}, let $\sX^c_{\bm{v}} = \R^n \setminus \sX_{\bm{v}}$ be its complement, which is open in $\R^n$.
We define the \emph{safety threshold value} $\Gamma^*:\R^l \to \R$
\begin{equation} \label{eq:safety-threshold}
  \Gamma^*(\bm{v}) \triangleq
    \inf_{\bm{x} \in\: \sX^c_{\bm{v}} \cap \sD_{\bm{v}}} V(\bm{x},\bm{v}) - \inf_{\bm{x} \in \sX_{\bm{v}} \cap \sD_{\bm{v}}} V(\bm{x}, \bm{v}).
\end{equation}
When positive, $\Gamma^*(\bm{v})$ represents the largest constraint-admissible level set of the Lyapunov function $V$ at reference $\bm{v}$. Our definition for $\Gamma^*$ is an extension of the \emph{threshold value} in \cite{nicotra2018explicit} to the entire reference space $\R^l$. With this, $\Gamma^*$ has the following property.

\begin{lemma}\label{lem:safety-threshold-iff}
  Let $V:\R^n \times \R^l \to \R$ be a reference-dependent Lyapunov function that satisfies Assumption \ref{ass:lyap-regular}. We have that $\Gamma^*(\bm{v}) \geq 0$ if and only if $\bm{v} \in \sV$.
\end{lemma}
\begin{proof}
  Let $\bm{v} \in \sV$ be given. Then, $\bar{x}(\bm{v}) \in \sX_{\bm{v}} \cap \sD_{\bm{v}}$, which implies that $\inf_{\bm{x} \in \sX_{\bm{v}} \cap \sD_{\bm{v}}} V(\bm{x},\bm{v}) = 0$. Recalling that $\forall \bm{x} \in \sD_{\bm{v}}$, $V(\bm{x},\bm{v}) \geq 0$, we conclude that $\Gamma^*(\bm{v})\geq 0$.

  For the other direction, we show the contrapositive statement. Let $\bm{v} \in \R^l \setminus \sV$ be given. Then, $\bar{x}(\bm{v}) \in \sX_{\bm{v}}^c \cap \sD_{\bm{v}}$, which implies that $\inf_{\bm{x} \in \sX_{\bm{v}}^c \cap \sD_{\bm{v}}} V(\bm{x},\bm{v}) = 0$. Since $\sX_{\bm{v}}^c \cap \sD_{\bm{v}}$ is open in $\R^n$, there exists $\delta > 0$ such that $\sB\big(\bar{x}(\bm{v}), \delta\big) = \{\bm{x} \in \R^n \mid \|\bm{x} - \bar{x}(\bm{v})\| < \delta\} \subset \sX_{\bm{v}}^c \cap \sD_{\bm{v}}$. Thus,
  \begin{equation}
      \Gamma^*(\bm{v}) = -\inf_{\bm{x} \in \sX_{\bm{v}} \cap \sD_{\bm{v}}} V(\bm{x}, \bm{v}) \leq -\gamma_1(\delta) < 0.
  \end{equation}
\end{proof}

Let us also define the \emph{stability threshold value} $\overline{\Gamma}: \R^l \to \R$
\begin{equation}\label{eq:stability-threshold}
  \overline{\Gamma}(\bm{v}) \triangleq  \inf_{\bm{x} \in \partial \sD_{\bm{v}}} V(\bm{x},\bm{v}).
\end{equation}
We can now construct DSMs using Lyapunov functions.

\begin{theorem}[\bf{Lyapunov-based DSMs}] \label{thm:lyap-dsm}
  Let $V:\R^n \times \R^l \to \R$ be a reference-dependent Lyapunov function for the prestabilized system $f_{\pi}$ satisfying Assumption \ref{ass:lyap-regular}. If $\Gamma^*: \R^l \to \R$ and $\overline{\Gamma}: \R^l \to \R$ are continuously differentiable, and $\sX \subset \R^n$ and $\sV \subset \R^l$ are compact, then for any $\epsilon \in (0,1)$,
  \begin{equation} \label{eq:lyap-dsm}
    \Delta(\bm{x}, \bm{v}) = \mat{\Gamma^*(\bm{v}) - V(\bm{x}, \bm{v}) \\ (1-\epsilon)\overline{\Gamma}(\bm{v}) - V(\bm{x}, \bm{v})},
  \end{equation}
  is a dynamic safety margin.
\end{theorem}
\begin{proof}
  To prove property \eqref{eq:dsm-safe}, let $(\bm{x}, \bm{v}) \in \tilde{\sC} \subset \tilde{\sD}$. Then, $\Gamma^*(\bm{v}) \geq V(\bm{x}, \bm{v}) \geq 0$, which implies that $\bm{v} \in \sV$ by Lemma \ref{lem:safety-threshold-iff}. Next, for a contradiction, assume $\bm{x} \notin \sX_{\bm{v}}$. Then
  \begin{equation*}
      \Gamma^*(\bm{v}) = \inf_{\bm{x} \in \sX_{\bm{v}}^{c}\cap \sD_{\bm{v}}} V(\bm{x}, \bm{v}) \leq V(\bm{x}, \bm{v}).
  \end{equation*}
  So, we must have $V(\bm{x}, \bm{v}) = \Gamma^*(\bm{v})$. If $V(\bm{x},\bm{v}) = 0$, then $\bm{x} = \bar{x}(\bm{v}) \in \sX_{\bm{v}}$ because $\bm{v} \in \sV$, and we have a contradiction.
  Otherwise, if $\Gamma^*(\bm{v}) = V(\bm{x},\bm{v}) > 0$, recall that $\sX^c_{\bm{v}} \cap \sD_{\bm{v}}$ is open in $\R^n$. Therefore, there exists an open set $\sN \subset \sX^c_{\bm{v}} \cap \sD_{\bm{v}}$ with $\bm{x} \in \sN$.
  Furthermore, since $\bm{x} \in \sD_{\bm{v}} \setminus \{\bar{x}(\bm{v})\}$ is a regular point of $V(\cdot, \bm{v})$ by Assumption \ref{ass:lyap-regular}, we have that $\frac{\partial V}{\partial \bm{x}} \neq 0$. So, there exists a sufficiently small scalar $\delta > 0$ such that $\hat{\bm{x}} = \bm{x} - \delta \frac{\partial V}{\partial \bm{x}}^\top  \in \sN \subset \sX^c_{\bm{v}} \cap \sD_{\bm{v}}$.
  Furthermore, this point achieves lower value in $V$. That is, $V(\hat{\bm{x}}, \bm{v}) < V(\bm{x}, \bm{v})$. So, $\Gamma^*(\bm{v}) = V(\bm{x},\bm{v})$ is a contradiction because
  \begin{equation*}
    \Gamma^*(\bm{v}) = \inf_{\bm{x} \in \sX^c_{\bm{v}} \cap \sD_{\bm{v}}} V(\bm{x},\bm{v}) \leq V(\hat{\bm{x}}, \bm{v}) < V(\bm{x}, \bm{v}).
  \end{equation*}
  
  To prove property \eqref{eq:dsm-compact}, note that since $\tilde{\sC} \subset \sX \times \sV$, and $\sX \times \sV$ is compact in $\R^n \times \R^l$, we must only show that $\tilde{\sC}$ is closed in $\R^n \times \R^l$. Let $\{(\bm{x}_k, \bm{v}_k)\} \subset \tilde{\sC}$ be a sequence that converges to $(\bm{x}, \bm{v}) \in \R^n \times \R^l$. Since $\sX \subset \R^n$ and $\sV \subset \R^l$ are closed, it follows that $\bm{x} \in \sX$ and $\bm{v} \in \sV$.
  Now we argue $\Delta(\bm{x}, \bm{v}) \geq 0$. For a contradiction, assume there is $i \in \{1,2\}$ such that $\Delta_i(\bm{x}, \bm{v}) < 0$. By continuity of $\Delta_i$ on $\R^n \times \R^l$, there exists an open neighborhood $\tilde{\sN} \subset \R^n \times \R^l$ of $(\bm{x}, \bm{v})$ wherein $\Delta_i(\bm{x}, \bm{v}) < 0$. By convergence of $\{(\bm{x}_k, \bm{v}_k)\}$ to $(\bm{x}, \bm{v})$, there exists $K \in \N$ such that $(\bm{x}_K, \bm{v}_K) \in \tilde{\sN}$, which raises a contradiction because $(\bm{x}_K, \bm{v}_K) \in \tilde{\sC}$ implies that $\Delta(\bm{x}_K, \bm{v}_K) \geq 0$.
  It remains to show that $(\bm{x}, \bm{v}) \in \tilde{\sD}$. Begin by noting that for all $k \in \N$, $(\bm{x}_k, \bm{v}_k) \in \tilde{\sD}$ implies that $V(\bm{x}_k, \bm{v}_k) \geq \gamma_1\big(\|\bm{x}_k - \bar{x}(\bm{v}_k)\|\big)$. It follows by continuity of all functions involved in the inequality and by the order limit theorem that $V(\bm{x}, \bm{v}) \geq \gamma_1\big(\|\bm{x} - \bar{x}(\bm{v})\|\big)$. For a contradiction, let us assume that $(\bm{x}, \bm{v}) \in \partial \tilde{\sD}$, which implies that $\bm{x} \in \partial \sD_{\bm{v}}$ and, therefore, $\overline{\Gamma}(\bm{v}) \leq V(\bm{x}, \bm{v})$. However, recall that we have shown $\Delta(\bm{x}, \bm{v}) \geq 0$. Thus, $V(\bm{x}, \bm{v})  \leq (1-\epsilon) \overline{\Gamma}(\bm{v})$. This yields
  \begin{equation*}
    \frac{1}{1-\epsilon} V(\bm{x}, \bm{v}) \leq \bar{\Gamma}(\bm{v})
    \leq V(\bm{x}, \bm{v}),
  \end{equation*}
  which is a contradiction because $V(\bm{x}, \bm{v}) \geq \gamma_1\big(\|\bm{x} - \bar{x}(\bm{v})\|\big) > 0$. So, we must have that $(\bm{x}, \bm{v}) \in \tilde{\sD}$ and we conclude that $\tilde{\sC}$ is closed in $\R^n \times \R^l$, and therefore compact.
  
  To prove property \eqref{eq:dsm-invariant}, let $i \in \{1,2\}$ be given and pick $(\bm{x}, \bm{v}) \in \tilde{\sD}$ such that $\Delta_i(\bm{x},\bm{v}) = 0$. Then, by \eqref{eq:lyap-decreases},
  \begin{equation}
    \frac{\partial \Delta_i}{\partial \bm{x}} f_{\pi}(\bm{x},\bm{v}) = -\frac{\partial
    V}{\partial \bm{x}} f_{\pi}(\bm{x},\bm{v}) \geq 0.
  \end{equation}
\end{proof}

The following corollary highlights that for Lyapunov-based DSMs, any $\alpha \in \cK$ can be used for the DSM-CBF policy $\tilde{\beta}$ given in \eqref{eq:beta-dsm}. This result constitutes a step away from existence-type proofs common in the CBF literature and towards implementation.
\begin{corollary} \label{cor:lyap-dsm-any-alpha-works}
  If $\Delta$ is a Lyapunov-based DSM as in \eqref{eq:lyap-dsm}, then $\forall \alpha \in \cK$, $\forall (\bm{x}, \bm{v}) \in \tilde{\sC}$, $\tilde{\sK}_{\alpha}(\bm{x}, \bm{v}) \neq \emptyset$.
\end{corollary}
\begin{proof}
Let $\alpha \in \cK$ and $(\bm{x}, \bm{v}) \in \tilde{\sC}$ be given. Recall that $\pi(\bm{x}, \bm{v}) \in \sU$ and note that for any $i \in \{1,2\}$,
\begin{equation*}
  \dot{\Delta}_i\big(\bm{x}, \bm{v}, \pi(\bm{x}, \bm{v}), 0\big) \!=\! -\frac{\partial V}{\partial \bm{x}} f_{\pi}(\bm{x}, \bm{v}) \!\geq\! 0 \!\geq\! - \alpha\big(\Delta_i(\bm{x}, \bm{v})\big).
\end{equation*}
Therefore, $\big(\pi(\bm{x}, \bm{v}), 0\big) \in \tilde{\sK}_{\alpha}(\bm{x},\bm{v})$.
\end{proof}

\begin{remark} \label{rem:multiple-thresholds}
  When multiple constraints are considered, it is usually simpler to define multiple safety threshold values $\Gamma^*_i$ and stack them into the form of \eqref{eq:lyap-dsm}. As long as each one is differentiable, Theorem \ref{thm:lyap-dsm} can still be applied to obtain a DSM $\Delta: \tilde{\sD} \to \R^{n_c}$ with $n_c > 2$. In the event either $\Gamma^*_i$ or $\overline{\Gamma}$ are not continuously differentiable, any continuously differentiable, lower-bounding function can be used in its place.
\end{remark}

\begin{remark} \label{rem:initial-reference}
  When choosing the initial virtual reference $\bm{v}$ in the augmented system \eqref{eq:sys-aug}, it is worth noting that any $\bm{v}_0 \in \sV$ such that $(\bm{x}_0,\bm{v}_0) \in \tilde{\sC}$ is valid. A desirable choice is
  \begin{equation}
      \bm{v}_0 = \underset{\bm{v} \in \sV}{\arg \min}~ V(\bm{x}_0, \bm{v}),
  \end{equation}
  because it tends to maximize the value of $\Delta$. If the above problem is infeasible, then $\bm{x}_{0}$ is not a safe starting condition for the designed DSM.
\end{remark}

\begin{remark} \label{rem:cbf->clf}
This section transforms the problem of finding a CBF into that of finding a Lyapunov function. The advantage in doing so is that the latter problem has a much richer history, which can be leveraged to synthesize Lyapunov-based DSMs using both analytical \cite{nicotra2015control,garone2018explicit, nicotra2018delay} and numerical \cite{cotorruelo2021reference} tools. It is also worth noting that one can always linearize the system around the current reference $v$ and use the Lyapunov function of the linearized system.
\end{remark}

\section{DSM-CBF Synthesis Approach Summary} \label{sec:summary}
In this manuscript, we have shown that dynamic safety margins (DSMs), key components in the explicit reference governor (ERG), are CBFs for an augmented system obtained by concatenating the state and reference vectors. This insight allows the direct use of approaches that compute DSMs to synthesize CBFs. Existing ERG works like \cite{nicotra2018explicit} outline different approaches to construct DSMs. This manuscript studied the Lyapunov-based approach, which requires finding a reference-dependent Lyapunov function \eqref{eq:lyap} for the prestabilized system and evaluating the safety \eqref{eq:safety-threshold} and stability \eqref{eq:stability-threshold} threshold values. These ingredients can then be combined to form a DSM. Once a DSM is obtained, a safe constrained control policy that minimally adjusts the nominal controller can be readily implemented by solving \eqref{eq:beta-dsm}. The following examples illustrate using the Lyapunov-based approach to construct DSM-CBFs and implementing \eqref{eq:beta-dsm}. Additional examples can be found in \cite{freire2025designing}, where the approach presented in this paper is specialized to Euler--Lagrange systems.

\section{Examples} \label{sec:examples}
In this section, we present two nonlinear examples to showcase the presented approach of designing CBFs using dynamic safety margins. For each example, we provide a comparison with candidate CBFs, backup CBFs, and the ERG. Simulations are computed in MATLAB and optimization problems are solved with MOSEK  \cite{mosek}.

\subsection{Anthill system}
\begin{figure}
    \centering
    \includegraphics[trim={0.4in 0 0.5in 0},width=0.9\linewidth]{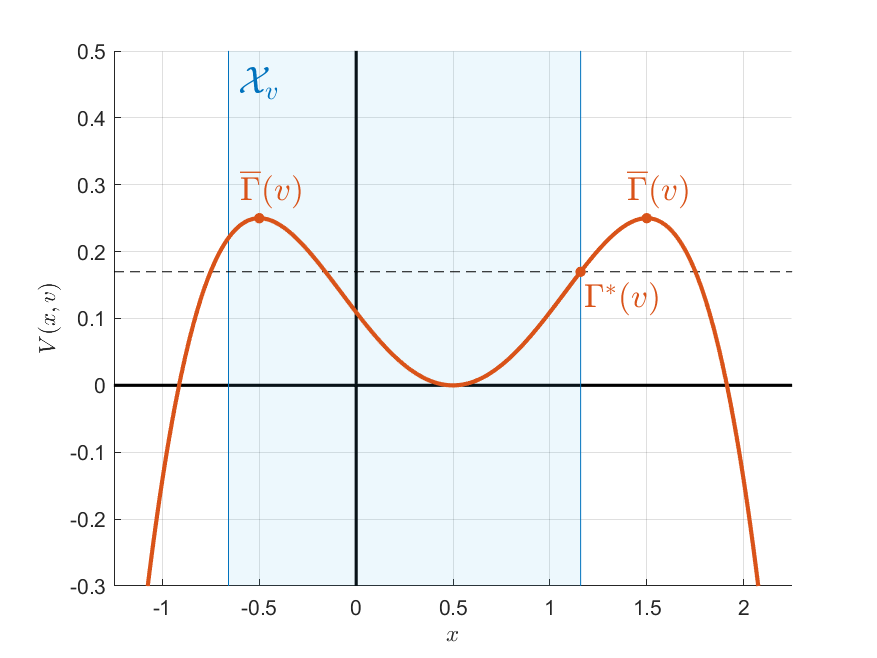}
    \caption{Anthill-shaped Lyapunov function $V(x,v)$ at $v = 0.5$, with the safety threshold value $\Gamma^*(v)$ and stability threshold value $\overline{\Gamma}(v)$.}
    \label{fig:anthill-lyap}
\end{figure}

Consider the scalar system
\begin{equation}\label{eq:anthill}
  \dot{x} = (x^2-1)x + u.
\end{equation}
In the absence of inputs $u \! = \!0$, the system asymptotically tends to $0$ if $|x_0| \!< \!1$ and diverges to infinity if $|x_0|>1$.
Let $\sU = [-u_{\max}, u_{\max}]$ be the input constraint
set, with $u_{\max} \!=\! \frac{4}{3\sqrt{3}}$.\\ Given the prestabilizing controller
\begin{equation}
  \pi(x,v) = -(v^2 - 1)v - 3xv(x-v),
\end{equation}
with $v \in \R$, the prestabilized system is
\begin{equation}
  \dot{x} = f_{\pi}(x,v) = (x-v)^3-(x-v),
\end{equation}
which satisfies the equilibrium conditions $\bar{x}(v) = v$ and $\bar{u}(v) = \pi\big(\bar{x}(v),v\big) = v - v^3$ for all $v \in \R$. The equilibrium point $\bar{x}(v)$ is attractive if $|x_0-v|<1$. This can be proven using a Lyapunov function constructed using the variable gradient method \cite{khalil2002nonlinear}
\begin{equation}
  V(x,v) = \int_{v}^x\!\!\!\! -f_{\pi}(\xi,v) \mathrm{d} \xi
  = \frac{1}{2}(x-v)^2-\frac{1}{4}(x-v)^4.
\end{equation}
This anthill-shaped function 
(see Fig. \ref{fig:anthill-lyap}) is such that, given the class $\cK$ functions $\gamma_1(c) = \frac{1}{4}c^2$ and $\gamma_2(c) = \frac{1}{2}c^2$, $\forall (x,v) \in \tilde{\sD} = \{(x,v) \in \R\times \R : x \in (v-1, v+1)\}$,
\begin{align*}
  \gamma_1(|x-v|) &\leq V(x,v) \leq \gamma_2(|x -v|), \\
  \frac{\partial V}{\partial x} f_{\pi}(x,v) &= -f_{\pi}^2(x,v) \leq 0,
\end{align*}
The function $V(x,v)$ also satisfies Assumption \ref{ass:lyap-regular}. The stability threshold value \eqref{eq:stability-threshold} is \begin{equation} \label{eq:anthill-gammabar}
\overline{\Gamma}(v)=V(v\pm1,v)=\frac14.
\end{equation}

To obtain the steady-state admissible set $\sV \subset \R$, consider the input constraints at steady-state $|\bar{u}(v)| = |v-v^3| \leq u_{\max}$. Using Cardano's formula to compute the roots of a 3-rd degree polynomial, we obtain $\sV = \left[-v_{\max}, v_{\max}\right]$, where
\begin{equation}
  v_{\max} = \frac{(2+\sqrt{3})^{\frac{1}{3}} + (2-\sqrt{3})^{\frac{1}{3}}}{\sqrt{3}}.
\end{equation}
The input constraints also induce state constraints for the prestabilized system. By setting $\pi(x,v) = \pm u_{\max}$ and solving for $x$, we obtain the reference-dependent state constraint set.
\begin{equation*}
  \sX_{v} = 
  \begin{cases}
    \emptyset, \quad & v < -4/\sqrt{3},\\
    \frac{1}{2}v+\left[\delta^-(v), -\delta^-(v)\right], \quad &
    v\in\left[-4/\sqrt{3},0\right),\\
    \R, \quad & v = 0, \\
    \frac{1}{2}v+\left[-\delta^+(v), \delta^+(v)\right], \quad & v\in\left(0,
    4/\sqrt{3}\right],\\
    \emptyset, \quad & v> 4/\sqrt{3},
  \end{cases}
\end{equation*}
where 
\begin{equation}
  \delta^{\pm}(v) = \frac{\sqrt{-3v^4 + 12v(v \pm u_{\max})}}{6v}.
\end{equation}
The resulting safety threshold value is 
\begin{equation*}
  \Gamma^*(v) \!=\! \begin{cases}
    -\infty, \;  &v
    \in (-\infty,\! -1\!-\!v_{\max}],\\
    -V\left(\frac{1}{2}v + \delta^-(v), v\right), \; &v \in(-1\!-\!v_{\max},\! -v_{\max}],\\
    V\left(\frac{1}{2}v + \delta^-(v),v\right), \; &v \in (-v_{\max}, 1 -v_{\max} ),\\
    \infty, \; &v \in [1 \!-\!v_{\max}, \!v_{\max} \!-\! 1],\\
    V\left(\frac{1}{2}v + \delta^+(v),v\right), \; & v\in (v_{\max} \!-\! 1,\!
    v_{\max}),\\
    -V\left(\frac{1}{2}v + \delta^+(v),v\right), \; &v \in [v_{\max},\! v_{\max} \!+\!  1),\\
    -\infty, \; &v
    \in [v_{\max} \!+\! 1,\!  \infty).
  \end{cases}
\end{equation*}
While this function is discontinuous, it is bounded by the smooth function $\Gamma_s^*(v) = \max \left( \min\big( \Gamma^*(v), \frac{1}{4}\big), -\frac{1}{4} \right)$. Thus, by Theorem \ref{thm:lyap-dsm}, $\Delta: \tilde{\sD} \to \R^2$ defined as
\begin{equation}
    \Delta(x,v) = \mat{\Gamma_s^*(v) - V(x,v)\\ 0.99\overline{\Gamma}(v) - V(x,v)},
\end{equation}
is a DSM and, by Theorem \ref{thm:dsm-cbf}, a CBF with respect to the augmented system.

We now consider the state constraints $\sX \!=\! [-x_{\max}, x_{\max}]$, with $x_{\max} = 2$. Since $x_{\max}>v_{\max}$ and \eqref{eq:anthill} is a first order system (no overshoot), these constraints have already been enforced by the restriction $v \in \sV$ induced by $\tilde{\sC}$. Thus, no further action is needed.

\begin{remark}
  It is interesting to note that, given $x_{\max}>v_{\max}$, the value of $x_{\max}$ plays no role in the DSM-based CBF. Indeed, the primary concern is to prevent instability by maintaining control authority over the system: any loss in control authority makes constraint violation inevitable, regardless of the specific value of $x_{\max}$.
\end{remark}

\begin{figure}
    \centering
    \includegraphics[width=0.95\linewidth]{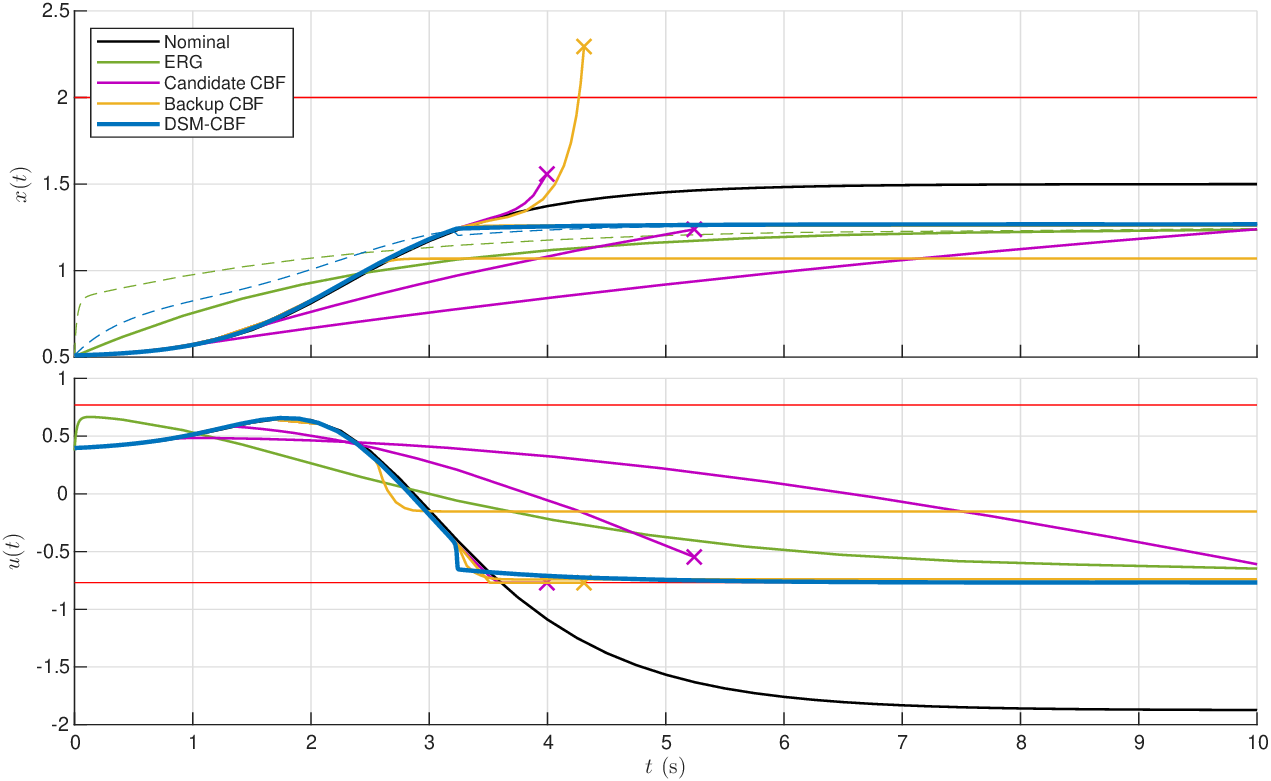}
    \caption{Closed-loop behavior for the anthill system using each of the considered constrained control approaches. The dashed green and blue lines represent the evolution of the virtual reference $v(t)$ for the ERG and the DSM-CBF approaches, respectively. In addition, we show the performance of the candidate CBFs for three different class $\cK$ functions $\alpha_1 = \alpha_2 = \alpha: c \mapsto a_i c$, where $a_i \in \{7, 0.15, 0.07\}$. Larger $a_i$ corresponds to better input tracking, but also an earlier failure due to infeasibility, marked by a magenta `$\times$' marker. Finally, we show the backup CBF performance for three different horizons $T \in \{0.1, 1, 5\}$. The short horizon results in the conservative behavior and the long horizon induces numerical issues that ultimately cause the approach to fail, marked with the `$\times$' symbol. The horizon $T= 1$ results in safe performance almost equal to our DSM-CBF.}
    \label{fig:anthill}
\end{figure}

Given the initial conditions $x_0 = 0.51$, $v_0 = 0.51$, the desired reference $r = 1.5$, and the nominal control law $\kappa(x,r)=\pi(x,r)$, we consider the following constrained control strategies: \begin{enumerate}[label=(\alph*)]
    \item DSM-CBF: We use $\eta = 0.01$, the class $\cK$ function $\alpha: c \mapsto 1.8c$, and $\rho:(v,r) \mapsto r-v$.
    \item ERG: We use $\rho_{\text{ERG}}:(v,r) \mapsto 100(r-v)$.
    \item Candidate CBF: We use the functions $h_1(x) = x_{\max} + x$ and $h_2(x) = x_{\max} - x$ and tune their class $\cK$ functions to illustrate the tradeoff between performance and feasibility that candidate CBFs suffer from.
    \item Backup CBF: We consider the backup policy 
    \begin{equation}
      \psi(x) = u_{\max} \left(\frac{1-e^{10x}}{1+e^{10x}}\right),
    \end{equation}
    which is a continuously differentiable approximation of $\psi^*(x) = -u_{\max}\text{sign}(x)$. The backup policy tries to drive the system to the origin using the maximum allowed input effort $u_{\max}$. Note that the set $\sS^0 = [-1, 1]$ is forward invariant under the backup dynamics $\dot{x} = f_{\psi}(x) \triangleq (x^2 - 1)x + \psi(x)$.
    Now, we can formulate the backup CBF
    \begin{equation*}
        h^{\psi}(x) \!=\! \min\left[ \min_{\tau \in [0,T]} h^{\sC}\big(\Phi_{f_{\psi}}(x, \tau)\big),
        h^{\sS}\big(\Phi_{f_{\psi}}(x,T)\big)\right],
    \end{equation*}
    where $T$ is the simulation horizon, $\Phi_{f_{\psi}}(x_0,\tau)$ is the solution of the backup dynamics $f_{\pi}$ at time $\tau$ with initial condition $x(0) = x_0$, $h^{\sC}(x) = x_{\max} - |x|$, and $h^{\sS}(x) = 1 - |x|$. The implementation is done following \cite{chen2021backup} with class $\cK$ function $\alpha: c \mapsto 7c$.
\end{enumerate}
The simulation results for each approach are shown in Fig. \ref{fig:anthill}. All approaches are able to enforce safety and reach the closest safe point to the desired reference but differ in performance and computational complexity. As expected, our DSM-CBF approach exhibits behavior that more closely resembles the nominal behavior than the ERG. With enough tuning, the candidate CBFs were able to also maintain safety for the fist 10 seconds, albeit with poor performance and no guarantee they will remain feasible afterwards. This is unsurprising since the candidate CBFs were constructed without any understanding of the inherent control authority constraint. The backup CBF approach with horizon $T = 1$ exhibits almost equal performance to our DSM-CBF approach. However, it must be noted that its computational footprint is significantly higher than our DSM-CBF approach due to the need at each time step to predict the backup trajectory forward in time. This computational footprint can be reduced by considering shorter time horizons (e.g. $T = 0.1$), but performance worsens as shown in Fig. \ref{fig:anthill}. The computational overhead of the safety filter associated with each approach is shown in Table \ref{tab:compute}. Moreover, the backup CBF approach is less numerically robust, as evidenced by the response to a longer time horizon $T = 10$. The reason the backup CBF fails for longer horizons is that it is numerically searching for the open boundary of the domain of attraction $x < v_{\max}$. Since a slight overshoot in the numerical integration leads to the divergence of the backup trajectory prediction, the backup CBF fails if it pushes too close to the boundary of the domain of attraction. Conversely, the DSM-based CBF avoids this issue by using the stability threshold $0.99\, \overline{\Gamma}(v)$, which yields a closed set with enough numerical leeway that small perturbations do not exit the domain of attraction of the prestabilizing controller. Our DSM-based CBF provided a systematic approach that successfully identified the constraints induced by the input saturation.

\begin{table}
    \centering
    \caption{Safety Filter Compute Times}
    \begin{tabular}{|c|c|c|c|c|}
      \hline
      & \multicolumn{2}{|c|}{\textbf{Anthill}} & \multicolumn{2}{|c|}{\textbf{Crane}}\\
      \hline
      \textbf{Approach} & \textbf{Ave} & \textbf{Max} & \textbf{Ave} & \textbf{Max}\\
      \hline
      Nominal & 0 & 0 & 0 & 0 \\
      \hline
      ERG   & 0 & 0 & 0 & 0\\
      \hline
      Candidate CBF & 3.46 & 4.27 & - & - \\
      \hline
      DSM-CBF (Proposed) & 3.54 & 4.76 & 3.64 & 6.94 \\
      \hline
      Backup CBF $T = 0.1$ & 4.51 & 5.33 & 7.85 & 15.26 \\
      \hline
      Backup CBF $T = 1.0$ & 4.99 & 6.11 & 12.98 & 22.93 \\
      \hline
      Backup CBF $T = 5.0$ & 10.92 & 17.25 & 20.39 & 35.97\\
      \hline
    \end{tabular}
    \label{tab:compute}
\end{table}

\subsection{Overhead Crane}
\begin{figure}
    \centering
    \includegraphics[width=0.6\linewidth]{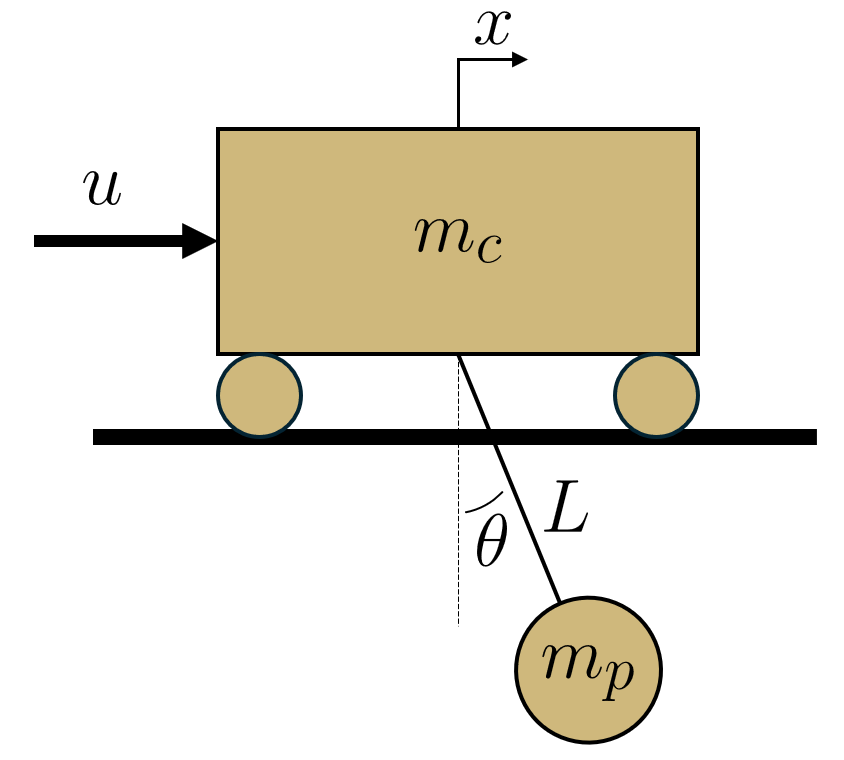}
    \caption{Overhead crane system from \cite{fang2001nonlinear}.}
    \label{fig:crane-model}
\end{figure}

Consider the dynamics of an overhead crane \cite{fang2001nonlinear}
\begin{equation}
    M(\bm{q})\ddot{\bm{q}} + C(\bm{q},\dot{\bm{q}})\dot{\bm{q}} + G(\bm{q}) = Bu,
\end{equation}
where the generalized coordinates $\bm{q} = [x; ~\theta]$ are the gantry position $x$ and the payload angle $\theta$, and
\begin{align*}
    M(\bm{q}) &= \mat{m_c + m_p & m_pL\cos\theta \\ m_pL\cos\theta & m_pL^2}, \quad B = \mat{1\\0},\\
   C(\bm{q},\dot{\bm{q}}) &= \mat{0 & -m_pL\dot{\theta}\sin\theta \\ 0 & 0}, \quad  G(\bm{q}) = \mat{0 \\ m_p gL\sin\theta},
\end{align*}
where $m_c, m_p > 0$ represent the gantry and payload masses, respectively, $L>0$ is the length of the rod connecting the crane and payload, and $g= 9.81$ is the acceleration of gravity (see Fig. \ref{fig:crane-model}).
It can be shown that, letting $\bm{x} = [\bm{q}; ~\dot{\bm{q}}]$, the system is control-affine and can be written as in \eqref{eq:sys}. We consider the prestabilizing PD control law given in \cite{fang2001nonlinear} \begin{equation} \label{eq:PDcontrol}
    \pi(\bm{x},v) = -k_p(x-v) - k_d\dot{x},
\end{equation} 
where $k_p > 0$, $k_d > 0$ are the proportional and derivative gains, and the equilibrium mapping is $\bar{x}:v \mapsto [v~;0~;0~;0]$. Since the overhead crane is an Euler--Lagrange system and $\pi$ is a passivity-based controller, let us formulate a storage function describing the total energy of the system \cite{ortega1998euler}
\begin{equation*}
    V(\bm{x},v) = \frac{1}{2}\dot{\bm{q}}^\top M(\bm{q}) \dot{\bm{q}} + m_pgL(1-\cos\theta) + \frac{1}{2}k_p(x-v)^2.
\end{equation*}
The storage function $V$ satisfies Definition $\ref{def:lyap}$ with $\sD_v = \R \times (-\pi, \pi) \times \R^2$ and also satisfies Assumption \ref{ass:lyap-regular}. We now define four types of constraints and, for each one, we find the associated safety threshold value or a continuously differentiable bounding function.
For details on the construction of each $\Gamma^*_i(v)$, the reader is referred to \cite{nicotra2015control,freire2025designing}. 
\begin{enumerate}
    \item Position constraints: $x_{\text{min}} \leq x \leq x_{\text{max}}$. For each of the two bounds, we take 
    \begin{align*}
        \Gamma_1^*(v) &=\frac{1}{2} k_p(v - x_{\text{min}})|v - x_{\text{min}}|, \\
        \Gamma_2^*(v) &= \frac{1}{2}k_p (x_{\text{max}} - v)|x_{\text{max}} - v|.
    \end{align*}
    \item Input constraints: $|u| \leq u_{\text{max}}$, with $u_{\text{max}} > 0$. Solving \eqref{eq:safety-threshold} yields the constant function
    \begin{equation*}
        \Gamma^*_3(v) = \frac{m_c u_{\text{max}}^2}{2(m_c k_p + k_d^2)}.
    \end{equation*}
    \item Angle constraints: $|\theta| \leq \theta_{\text{max}}$, with $\theta_{\text{max}} \in (0,\pi/2)$. Solving \eqref{eq:safety-threshold} yields the constant function
    \begin{equation*}
        \Gamma_4^*(v) = m_pgL(1-\cos \theta_{\text{max}}).
    \end{equation*}

    \item Payload constraints: $x + L\sin\theta \leq p_{\text{max}}$, with $p_{\text{max}} \geq x_{\text{max}}$. For this constraint, we were unable to find an analytical expression for $\Gamma_5^*(v)$ associated to $V$. However, as detailed in \cite[Proposition 4]{nicotra2015control}, we can define
    \begin{equation*}
        \underline{V}(\bm{x},v) = \frac{1}{2}\dot{\bm{q}}^\top M(\bm{x}) \dot{\bm{q}} + \frac{4}{\pi^2}m_pgL\theta^2 + \frac{1}{2}k_p(x-v)^2,
    \end{equation*}
    which satisfies $\underline{V}(\bm{x},v) \leq V(\bm{x},v)$, $\forall \theta \in (-\pi/2, \pi/2)$. Moreover, since $x_{\text{max}}\leq p_{\text{max}}$, we can perform a planar embedding of the payload constraint
    \begin{equation}
        x + L\theta \leq p_{\text{max}} \quad \implies \quad x + L \sin\theta \leq p_{\text{max}},
    \end{equation}
    which holds $\forall x\leq x_{\text{max}}$ and $\forall \theta\in[-\pi/2,\pi/2]$. With these approximations, the following continuously differentiable function is a bound for $\Gamma^*_5(v)$ and can be used in its stead:
    \begin{equation*}
        \Gamma_5(v) = \frac{4k_pm_pg}{8m_p g + Lk_p\pi^2} (p_{\text{max}} - v)|p_{\text{max}} - v|.
    \end{equation*}
\end{enumerate}
The stability threshold value is $\overline\Gamma(v)=2m_pgL$.
Using Theorem \ref{thm:lyap-dsm}, we conclude that 
\begin{equation*}
    \Delta(\bm{x},v) = \mat{\Gamma^*_1(v) - V(\bm{x}, v)\\ \Gamma^*_2(v) - V(\bm{x}, v)\\
    \Gamma^*_3(v) - V(\bm{x}, v)\\
    \Gamma^*_4(v) - V(\bm{x}, v)\\
    \Gamma_5(v) - V(\bm{x}, v)\\
    0.99\overline{\Gamma}(v) - V(\bm{x}, v)},
\end{equation*}
is a DSM and, by Theorem \ref{thm:dsm-cbf}, a CBF with respect to the augmented system. Note that the stability threshold value satisfies $\Gamma^*_4(v) < \overline{\Gamma}(v)$. As such, when the angle constraints are enforced, there is no need to include $\overline{\Gamma}(v)$ in the definition of $\Delta$ above.
    
For all simulations, we use linear class $\cK$ functions given by $\alpha_i: c \mapsto \alpha_i c$ where $\alpha_i > 0$ represents both the function and the scalar gain despite the slight abuse of notation. We choose $\alpha_i = 100$ for all $i \in \{1,\ldots,6\}$. The navigation field we use is $\rho(v,r) = r-v$.

For comparison, we consider the following approaches.
\begin{enumerate}[label=(\alph*)]
    \item Nominal: The nominal controller is a PD control law $\kappa(\bm{x},r) = -k_{p,\kappa} (x - r) - k_{d,\kappa} \dot{x}$, where the gains $k_{p,\kappa}> 0$ and $k_{d,\kappa} > 0$ can differ from the ones in \eqref{eq:PDcontrol} and $r \in \R$ is the target reference.
    \item ERG: The explicit reference governor uses the same dynamic safety margins as our approach, but changing $k_p \to k_{p,\kappa}$ and $k_d \to  k_{d,\kappa}$ in all the expressions. The navigation field is $\rho_{\text{ERG}}(v,r) = 1000(r - v)$. The ERG-based, closed-loop system is
    \begin{equation}
        \mat{\dot{\bm{x}} \\ \dot{v}} = \mat{f(\bm{x}) + g(\bm{x}) \kappa(\bm{x},v) \\ 
        \min_{i} \Delta_i(\bm{x},v) \rho_{\text{ERG}}(v,r)}.
    \end{equation}
     
    \item Backup CBF: We consider the backup policy
    \begin{equation}
      \psi(\bm{x}) = u_{\max} \tanh\left(\frac{-K_{\psi}\bm{x}}{u_{\max}}\right),
    \end{equation}
    where $K_{\psi} = [3.15 ~ -16.35 ~ 5.48 ~ -5.16]$ was obtained linearizing the system around the origin and solving the LQR problem. The $\tanh$ term is a smooth clipping function that ensures $|\psi(\bm{x})| \leq u_{\max}$. An invariant set associated with this backup policy is $\sS^0 = \{\bm{x} \in \R^4 \mid \bm{x}^\top P_{\psi} \bm{x} \leq \Gamma_{\psi}\}$, where $P_{\psi}$ is the symmetric, positive definite matrix that solves the Riccati equation associated with the LQR problem, and $\Gamma_{\psi} = 2.5$ is determined heuristically through trial and error in the simulation. We formulate the backup CBF
    \begin{equation*}
        h^{\psi}(\bm{x}) \!=\! \min\left[ \min_{\tau \in [0,T]} h^{\sC}\big(\Phi_{f_{\psi}}(\bm{x},\! \tau)\big),
        h^{\sS}\big(\Phi_{f_{\psi}}(\bm{x},\!T)\big)\right],
    \end{equation*}
    where $T$ is the simulation horizon, $\Phi_{f_{\psi}}(\bm{x}_0,\tau)$ is the solution of the backup dynamics at time $\tau$ with initial condition $\bm{x}(0) = \bm{x}_0$,
    \begin{equation}
      h^{\sC}(\bm{x}) = \min \mat{x_{\max} - |x|\\ \theta_{\max} - |\theta|\\  p_{\max} - x - L\sin\theta},
    \end{equation}
    and $h^{\sS}(\bm{x}) = \Gamma_{\psi} - \bm{x}^\top P_{\psi} \bm{x}$. The class $\cK$ functions used are $\alpha(c) = 40c$ for the position and payload constraints, $\alpha(c) = 100c$ for the angle constraints and $\alpha(c) = 400c$ for the terminal constraint $h^{\sS}$.
\end{enumerate}

\begin{figure}
    \centering
    \includegraphics[width=0.95\linewidth]{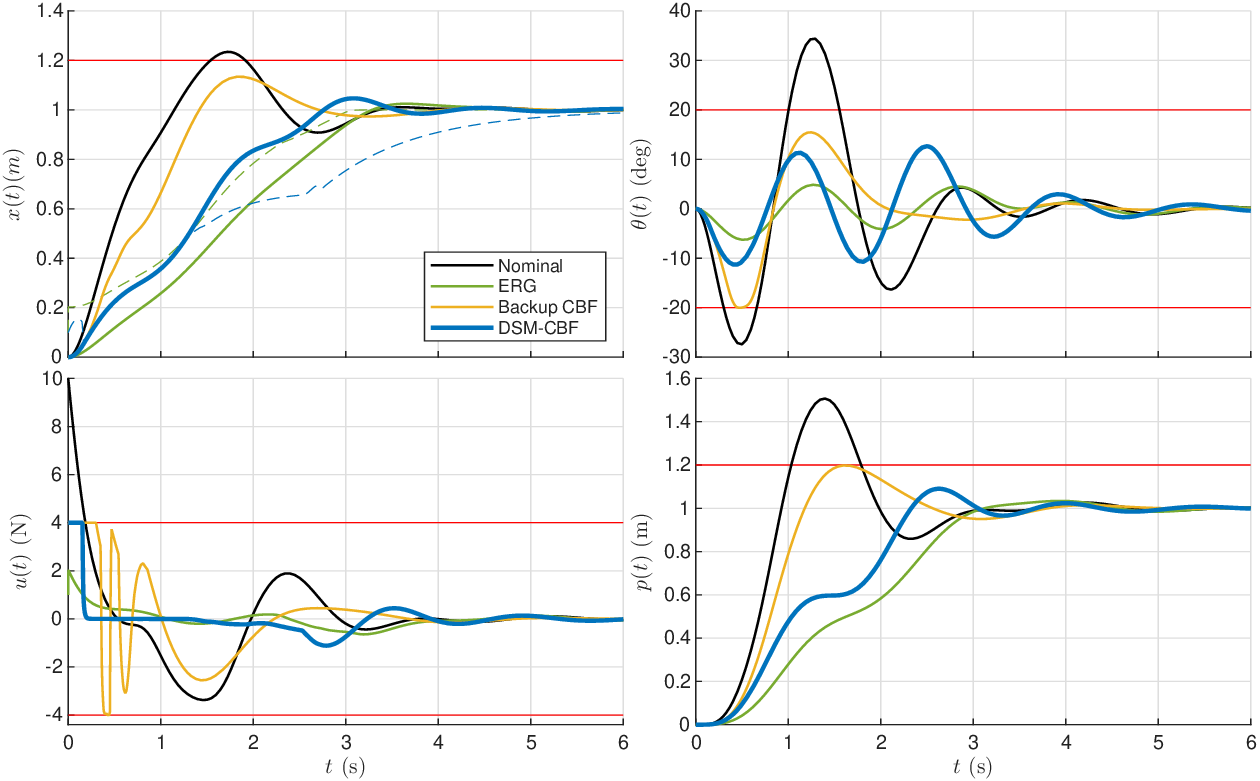}
    \caption{Closed-loop behavior of the overhead crane system under each of the considered constrained control approaches. The dashed green and blue lines represent the evolution of the virtual reference $v(t)$ for the ERG and DSM-CBF approaches, respectively. While only the trace for horizon $T = 5$ for the backup CBF approach is shown, the computational burden of other horizon choices is shown in Table \ref{tab:compute}.}
    \label{fig:crane}
\end{figure}

\begin{table}[b]
  \centering
  \caption{Overhead Crane Simulation Parameters}
  \begin{tabular}{|c|c|c|c|c|c|c|c|}
  \hline
  \!$m_c$ \!(\unit{kg})\! &\! $1$ \!& \!$k_p$\! &\! $2$ \!& \!$x_{\text{min}}$ \!(\unit{m})\! &\! $-1.2$ \\
  \hline
  \!$m_p$ \!(\unit{kg})\! &\! $0.5$ \!& \!$k_d$\! & $0.1$ \!&\! $x_{\text{max}}$ \!(\unit{m}) \!&\! $1.2$ \\
  \hline
  \!$L$ \!(\unit{m})\! &\! $0.7$ \!&\! \!$k_{p,\kappa}$\! &\! $10$ \!&\! $u_{\text{max}}$ \!(\unit{N}) \!&\! $4$\\
  \hline
  \!$\eta$ \!(-)\! &\! $0.1$ \!& \!$k_{d,\kappa}$\! &\! $4$ \!&\! $\theta_{\text{max}}$ \!(\unit{deg}) \!&\! $20$ \\
  \hline
  \! $r$ \! (\unit{m})\! &\! $1$ \!&\! $v(0)$ \! (\unit{m}) \! & \! $0.1$ \!&\! $p_{\max}$ \! (\unit{m})\! & \! $1.2$ \\
  \hline
  \end{tabular}
  \label{tab:sim-params}
\end{table}
General parameters for the example are provided in Table \ref{tab:sim-params}, the performance of each approach is shown in Figure \ref{fig:crane}, and the computational burden of each approach is shown in Table \ref{tab:compute}. While all approaches achieve the goal and respect the safety constraints, their performance differs significantly. The ERG exhibits the slowest response but has the least computational burden. On the other hand, the backup CBF approach tracks the nominal behavior closely at the expense of high computational cost. Our DSM-CBF approach sits in the middle with a strict performance improvement over the ERG at a modest computational burden.

\section{Conclusion} \label{sec:conclusion}
This paper showed that dynamic safety margins are control barrier functions for an augmented system obtained by concatenating the state and reference vectors. The work presents an optimization-based safe policy and it is shown that it is persistently feasible. Then, a Lyapunov-based approach to construct DSMs was presented in detail and its performance against other common approaches to design CBFs, as well as against the ERG was compared in two nonlinear systems.

Future work includes continuing to explore connections between the CBF and reference governor literatures, study how different Lyapunov functions lead to different CBFs and their impact on closed-loop performance, adapt other types of DSMs to construct CBFs, and robustifying the presented results to account for bounded disturbances.

\bibliographystyle{IEEEtran}
\bibliography{references}

\begin{thebibliography}{10}
\providecommand{\url}[1]{#1}
\csname url@samestyle\endcsname
\providecommand{\newblock}{\relax}
\providecommand{\bibinfo}[2]{#2}
\providecommand{\BIBentrySTDinterwordspacing}{\spaceskip=0pt\relax}
\providecommand{\BIBentryALTinterwordstretchfactor}{4}
\providecommand{\BIBentryALTinterwordspacing}{\spaceskip=\fontdimen2\font plus
\BIBentryALTinterwordstretchfactor\fontdimen3\font minus \fontdimen4\font\relax}
\providecommand{\BIBforeignlanguage}[2]{{%
\expandafter\ifx\csname l@#1\endcsname\relax
\typeout{** WARNING: IEEEtran.bst: No hyphenation pattern has been}%
\typeout{** loaded for the language `#1'. Using the pattern for}%
\typeout{** the default language instead.}%
\else
\language=\csname l@#1\endcsname
\fi
#2}}
\providecommand{\BIBdecl}{\relax}
\BIBdecl

\bibitem{molnar2023safety}
T.~G. Molnar, G.~Orosz, and A.~D. Ames, ``On the safety of connected cruise control: analysis and synthesis with control barrier functions,'' in \emph{Proc. IEEE Conf. Decis. Control (CDC)}, 2023, pp. 1106--1111.

\bibitem{heshmati2024control}
S.~Heshmati-Alamdari, M.~Sharifi, G.~C. Karras, and G.~K. Fourlas, ``Control barrier function based visual servoing for mobile manipulator systems under functional limitations,'' \emph{IAS Robot. Auton. Syst.}, p. 104813, 2024.

\bibitem{tayal2024control}
M.~Tayal, R.~Singh, J.~Keshavan, and S.~Kolathaya, ``Control barrier functions in dynamic {UAV}s for kinematic obstacle avoidance: A collision cone approach,'' in \emph{Proc. IEEE Amer. Control Conf. (ACC)}, 2024, pp. 3722--3727.

\bibitem{garone2017reference}
E.~Garone, S.~Di~Cairano, and I.~Kolmanovsky, ``Reference and command governors for systems with constraints: A survey on theory and applications,'' \emph{Automatica}, vol.~75, pp. 306--328, 2017.

\bibitem{ames2019control}
A.~D. Ames, S.~Coogan, M.~Egerstedt, G.~Notomista, K.~Sreenath, and P.~Tabuada, ``Control barrier functions: Theory and applications,'' in \emph{Proc. IEEE Eur. Control Conf. (ECC)}, 2019, pp. 3420--3431.

\bibitem{ames2016control}
A.~D. Ames, X.~Xu, J.~W. Grizzle, and P.~Tabuada, ``Control barrier function based quadratic programs for safety critical systems,'' \emph{IEEE Trans. Autom. Control (TAC)}, vol.~62, no.~8, pp. 3861--3876, Aug. 2017.

\bibitem{nguyen2016exponential}
Q.~Nguyen and K.~Sreenath, ``Exponential control barrier functions for enforcing high relative-degree safety-critical constraints,'' in \emph{Proc. IEEE Amer. Control Conf. (ACC)}, 2016, pp. 322--328.

\bibitem{breeden2021high}
J.~Breeden and D.~Panagou, ``High relative degree control barrier functions under input constraints,'' in \emph{Proc. IEEE Conf. Decis. Control (CDC)}, 2021, pp. 6119--6124.

\bibitem{xu2018constrained}
X.~Xu, ``Constrained control of input--output linearizable systems using control sharing barrier functions,'' \emph{Automatica}, vol.~87, pp. 195--201, Jan. 2018.

\bibitem{agrawal2017discrete}
A.~Agrawal and K.~Sreenath, ``Discrete control barrier functions for safety-critical control of discrete systems with application to bipedal robot navigation.'' in \emph{Proc. Robot.: Sci. Syst. (RSS)}, vol.~13, 2017, pp. 1--10.

\bibitem{zhang2022control}
Y.~Zhang, S.~Walters, and X.~Xu, ``Control barrier function meets interval analysis: Safety-critical control with measurement and actuation uncertainties,'' in \emph{Proc. IEEE Amer. Control Conf. (ACC)}, 2022, pp. 3814--3819.

\bibitem{jankovic2018robust}
M.~Jankovic, ``Robust control barrier functions for constrained stabilization of nonlinear systems,'' \emph{Automatica}, vol.~96, pp. 359--367, 2018.

\bibitem{buch2021robust}
J.~Buch, S.-C. Liao, and P.~Seiler, ``Robust control barrier functions with sector-bounded uncertainties,'' \emph{IEEE Control Syst. Lett. (LCSS)}, vol.~6, pp. 1994--1999, 2021.

\bibitem{clark2024semi}
A.~Clark, ``A semi-algebraic framework for verification and synthesis of control barrier functions,'' \emph{IEEE Trans. Autom. Control (TAC)}, 2024.

\bibitem{dai2023convex}
H.~Dai and F.~Permenter, ``Convex synthesis and verification of control-{L}yapunov and barrier functions with input constraints,'' in \emph{Proc. IEEE Amer. Control Conf. (ACC)}, 2023, pp. 4116--4123.

\bibitem{zeng2021decay}
J.~Zeng, B.~Zhang, Z.~Li, and K.~Sreenath, ``Safety-critical control using optimal-decay control barrier function with guaranteed point-wise feasibility,'' in \emph{Proc. IEEE Amer. Control Conf. (ACC)}, 2021, pp. 3856--3863.

\bibitem{choi2021robust}
J.~J. Choi, D.~Lee, K.~Sreenath, C.~J. Tomlin, and S.~L. Herbert, ``Robust control barrier--value functions for safety-critical control,'' in \emph{Proc. IEEE Conf. Decis. Control (CDC)}, 2021, pp. 6814--6821.

\bibitem{jang2024safe}
I.~Jang and H.~J. Kim, ``Safe control for navigation in cluttered space using multiple {L}yapunov-based control barrier functions,'' \emph{IEEE Robot. Autom. Lett. (RA-L)}, vol.~9, no.~3, pp. 2056--2063, 2024.

\bibitem{cortez2022safe}
W.~S. Cortez and D.~V. Dimarogonas, ``Safe-by-design control for {E}uler--{L}agrange systems,'' \emph{Automatica}, vol. 146, p. 110620, Dec. 2022.

\bibitem{saveriano2019learning}
M.~Saveriano and D.~Lee, ``Learning barrier functions for constrained motion planning with dynamical systems,'' in \emph{Proc. IEEE/RSJ Int. Conf. Intell. Robot. Syst. (IROS)}, 2019, pp. 112--119.

\bibitem{dai2022learning}
B.~Dai, P.~Krishnamurthy, and F.~Khorrami, ``Learning a better control barrier function,'' in \emph{Proc. IEEE Conf. Decis. Control (CDC)}, 2022, pp. 945--950.

\bibitem{chen2021backup}
Y.~Chen, M.~Jankovic, M.~Santillo, and A.~D. Ames, ``Backup control barrier functions: Formulation and comparative study,'' in \emph{Proc. IEEE Conf. Decis. Control (CDC)}, 2021, pp. 6835--6841.

\bibitem{freire2023systematic}
V.~Freire and M.~M. Nicotra, ``Systematic design of discrete-time control barrier functions using maximal output admissible sets,'' \emph{IEEE Control Syst. Lett. (LCSS)}, vol.~7, pp. 1891--1896, 2023.

\bibitem{nicotra2018explicit}
M.~M. Nicotra and E.~Garone, ``The explicit reference governor: A general framework for the closed-form control of constrained nonlinear systems,'' \emph{IEEE Control Syst. Mag.}, vol.~38, no.~4, pp. 89--107, Aug. 2018.

\bibitem{li2023governor}
Z.~Li and N.~Atanasov, ``Governor-parameterized barrier function for safe output tracking with locally sensed constraints,'' \emph{Automatica}, vol. 152, p. 110996, 2023.

\bibitem{bemporad1997nonlinear}
A.~Bemporad, A.~Casavola, and E.~Mosca, ``Nonlinear control of constrained linear systems via predictive reference management,'' \emph{IEEE Trans. Autom. Control (TAC)}, vol.~42, no.~3, pp. 340--349, 1997.

\bibitem{blanchini1999set}
F.~Blanchini, ``Set invariance in control,'' \emph{Automatica}, vol.~35, no.~11, pp. 1747--1767, 1999.

\bibitem{aubin1991viability}
J.~Aubin, \emph{Viability theory}.\hskip 1em plus 0.5em minus 0.4em\relax Birh{\"a}user, Boston, 1991.

\bibitem{nagumo1942lage}
M.~Nagumo, ``{\"U}ber die lage der integralkurven gew{\"o}hnlicher differentialgleichungen,'' in \emph{Proc. Phys.-Math. Soc. Jpn.}, vol.~24, 1942, pp. 551--559.

\bibitem{aubin2009set}
J.-P. Aubin and H.~Frankowska, \emph{Set-Valued Analysis}.\hskip 1em plus 0.5em minus 0.4em\relax Springer, 2009.

\bibitem{mangasarian1967fritz}
O.~L. Mangasarian and S.~Fromovitz, ``The {F}ritz {J}ohn necessary optimality conditions in the presence of equality and inequality constraints,'' \emph{J. Math. Anal. Appl.}, vol.~17, no.~1, pp. 37--47, 1967.

\bibitem{berge1963topological}
C.~Berge, \emph{Topological spaces: Including a treatment of multi-valued functions, vector spaces and convexity}.\hskip 1em plus 0.5em minus 0.4em\relax Oliver \& Boyd, 1963.

\bibitem{nicotra2015control}
M.~M. Nicotra and E.~Garone, ``Control of {Euler-Lagrange} systems subject to constraints: An explicit reference governor approach,'' in \emph{Proc. IEEE Conf. Decis. Control (CDC)}, Dec. 2015, pp. 1154--1159.

\bibitem{garone2018explicit}
E.~Garone, M.~Nicotra, and L.~Ntogramatzidis, ``Explicit reference governor for linear systems,'' \emph{Int. J. Control}, vol.~91, no.~6, pp. 1415--1430, 2018.

\bibitem{nicotra2018delay}
M.~M. Nicotra, T.~W. Nguyen, E.~Garone, and I.~V. Kolmanovsky, ``Explicit reference governor for the constrained control of linear time-delay systems,'' \emph{IEEE Trans. Autom. Control (TAC)}, vol.~64, no.~7, pp. 2883--2889, Jul. 2019.

\bibitem{li2020fast}
Z.~Li, {\"O}.~Arslan, and N.~Atanasov, ``Fast and safe path-following control using a state-dependent directional metric,'' in \emph{Proc. IEEE Int. Conf. Robot. Autom. (ICRA)}, 2020, pp. 6176--6182.

\bibitem{cotorruelo2021reference}
A.~Cotorruelo, M.~Hosseinzadeh, D.~R. Ramirez, D.~Limon, and E.~Garone, ``Reference dependent invariant sets: Sum of squares based computation and applications in constrained control,'' \emph{Automatica}, vol. 129, p. 109614, 2021.

\bibitem{khalil2002nonlinear}
H.~K. Khalil, \emph{{Nonlinear Systems}}.\hskip 1em plus 0.5em minus 0.4em\relax Upper Saddle River, NJ: Prentice-Hall, 2002.

\bibitem{rimon1991construction}
E.~Rimon and D.~E. Koditschek, ``The construction of analytic diffeomorphisms for exact robot navigation on star worlds,'' in \emph{Proc. IEEE Int. Conf. Robot. Autom. (ICRA)}, May 1989, pp. 21--26.

\bibitem{morris2013sufficient}
B.~Morris, M.~J. Powell, and A.~D. Ames, ``Sufficient conditions for the lipschitz continuity of qp-based multi-objective control of humanoid robots,'' \emph{Proc. IEEE Conf. Decis. Control (CDC)}, pp. 2920--2926, 2013.

\bibitem{mestres2025regularity}
P.~Mestres, A.~Allibhoy, and J.~Cort{\'e}s, ``Regularity properties of optimization-based controllers,'' \emph{Eur. J. Control}, vol.~81, p. 101098, 2025.

\bibitem{freire2026designing}
V.~Freire and M.~M. Nicotra, ``Designing control barrier functions using a dynamic backup policy,'' \emph{arXiv preprint arXiv:2510.09810}, 2026.

\bibitem{freire2025designing}
V.~Freire, S.~Debarshi, and M.~M. Nicotra, ``Designing control barrier functions for underactuated {E}uler--{L}agrange systems using dynamic safety margins,'' \emph{IEEE Control Syst. Lett. (LCSS)}, 2025.

\bibitem{mosek}
M.~ApS, \emph{The MOSEK optimization toolbox for MATLAB manual. Version 10.1.28.}, Copenhagen, Denmark, 2024.

\bibitem{fang2001nonlinear}
Y.~Fang, E.~Zergeroglu, W.~Dixon, and D.~Dawson, ``Nonlinear coupling control laws for an overhead crane system,'' in \emph{Proc. IEEE Int. Conf. Control Appl. (CCA)}, Sep. 2001, pp. 639--644.

\bibitem{ortega1998euler}
R.~Ortega, A.~Loria, P.~J. Nicklasson, and H.~Sira-Ramirez, \emph{{E}uler--{L}agrange systems}.\hskip 1em plus 0.5em minus 0.4em\relax Springer, 1998.

\end{thebibliography}

\end{document}